\documentclass[fleqn,usenatbib]{rasti}

\usepackage{newtxtext,newtxmath}

\usepackage[T1]{fontenc}

\DeclareRobustCommand{\VAN}[3]{#2}
\let\VANthebibliography\thebibliography
\def\thebibliography{\DeclareRobustCommand{\VAN}[3]{##3}\VANthebibliography}

\usepackage{graphicx}	
\usepackage{multirow}
\usepackage{amsmath}	
\usepackage{subfig}
\usepackage{comment} 
\usepackage{tikz} 
\usetikzlibrary{shapes.geometric,shapes.misc,arrows,arrows.meta,backgrounds,fit,positioning,calc}
\tikzstyle{startstop} = [rectangle, thick,rounded corners, minimum width=3cm, minimum height=1cm,text centered, draw=black, fill=red!30]
\tikzstyle{process} = [rectangle,thick, minimum width=3cm, minimum height=1cm, text centered, draw=black, fill=orange!30]
\tikzstyle{decision} = [rectangle, thick, minimum width=3cm, minimum height=1cm, text centered, draw=black, fill=green!30]
\tikzstyle{arrow} = [very thick,->,>=stealth]

\title[PHOTO-CAT - Photometric Contamination Analyzer Tool]{PHOTO-CAT - Photometric Contamination Analyzer Tool: A High-Performance Framework for Photometric Contamination Assessment}

\author[G. Galletta et al.]{G. Galletta$^{1,2,3}$\thanks{Contact for questions related to the PHOTO-CAT tool: gabriele.galletta@inaf.it},
F. Favata$^{3,4,5}$, L. Bradley$^{4}$, A. Saba$^{6,4}$\thanks{arianna.1.saba@kcl.ac.uk}, R. Grant$^{4}$, J. McLaren$^{4}$, S. Harvey$^{4}$, \newauthor
M. Tessenyi$^{4}$ and G. Micela$^{3}$
\\
$^{1}$Dipartimento di Fisica e  Chimica, Università di Palermo, Via Archirafi 36, 90128, Palermo, Italy.\\
$^{2}$Blue Skies Space Italia S.R.L. Via Vincenzo Monti 16, 20123, Milano, Italy.\\
$^{3}$INAF, Osservatorio Astronomico di Palermo, Palermo, Italy.\\
$^{4}$Blue Skies Space Ltd., 69 Wilson Street, London, EC2A 2BB, UK.\\
$^{5}$Department of Physics, Imperial College London, Exhibition Road, London SW7 2AZ, United Kingdom\\
$^{6}$Department of Physics, King’s College London, Strand, London WC2R 2LS, UK
}

\date{Accepted 2026 November 20. Received 2026 October 28; in original form 2026 June 10}

\pubyear{\the\year{}}

\begin{document}
\label{firstpage}
\pagerange{\pageref{firstpage}--\pageref{lastpage}}
\maketitle

\begin{abstract}
High-precision photometric missions are sensitive to flux contamination from neighboring stars, which can bias light curves, dilute transit depths, and compromise photometric precision. We present PHOTO-CAT, a catalog-driven, mission-agnostic tool for quantifying stellar flux contamination in photometric observations. The tool builds a three-dimensional Euclidean representation of the sky from a user-provided astrometric catalog, indexes it with a memory-optimized KDTree for rapid neighbor searches, and computes per-target contamination metrics from angular separation, relative fluxes, and configurable aperture models (top-hat or circular Gaussian PSF), optionally converted into an arbitrary instrumental bandpass via empirical photometric relations or stellar spectral models. A streaming, CSR-style storage format with compact internal identifiers enables scalable processing of catalogs with tens to hundreds of millions of sources with minimal memory overhead. Outputs include per-target contaminating flux, contamination ratios, target purity, number of neighboring sources, and their angular separations, supporting target screening and field-selection decisions ahead of observations. We demonstrate the software performance on all-sky Gaia samples (G $\leq$ 12, with neighbors down to G = 17; over three million targets), providing global contamination statistics and spatial characterization of crowded regions, and apply it to the Ariel Mission Candidate Sample to show how wavelength-dependent contamination estimates support target selection and observing-geometry optimization. Although developed for the Mauve satellite, PHOTO-CAT is mission-agnostic and readily applicable to TESS, PLATO, Ariel, and other surveys.
\end{abstract}

\begin{keywords}
Algorithms -- star -- contamination -- catalog
\end{keywords}



\section{Introduction}\label{sec:introduction}

The pursuit of precise stellar and exoplanetary characterization has driven photometric observations toward increasingly stringent requirements on measurement fidelity. Space-based observatories such as Kepler \citep[][]{borucki2010kepler, koch2010kepler}, TESS \citep[][]{ricker2016transiting}, and CHEOPS \citep[][]{benz2021cheops} have demonstrated the transformative potential of high-precision photometry in detecting exoplanets and studying stellar variability, while forthcoming missions such as PLATO \citep[][]{rauer2014plato, rauer2025plato} and Ariel \citep[][]{tinetti2016science, tinetti2018chemical, 2022EPSC...16.1114T} are expected to further expand these capabilities. This precision can be compromised by photometric contamination, i.e. the blending of flux from neighboring stars within a telescope's point-spread function (PSF) or photometric aperture, which introduces systematic biases in measured light curves and can lead to false positives, erroneous planetary parameters, or inaccurate stellar variability assessments \citep[][]{2024CoSka..54b.142D, nascimbeni2022plato}.\\
The impact of flux contamination on transit photometry was already recognized in the Kepler era, especially in connection with unresolved stellar companions. \cite{2015ApJ...805...16C} showed that neglecting stellar multiplicity leads to systematically underestimated planetary radii in transit surveys such as Kepler, K2, and TESS, and \cite{2020ApJ...898...47F} showed that fitting unresolved binary hosts as single stars also biases the derived stellar parameters. \cite{2022ApJ...935..141S} and \cite{2022AJ....164..138S} developed a framework to revise the properties of planet-host binary systems, finding that apparent near-Earth-analog planets in binaries are often sub-Neptunes once the flux dilution by the companion is accounted for, while \cite{2021AJ....162...75L} used speckle observations of TESS hosts to detect stellar companions and quantify their implications for small-planet detection.\\
The problem of contamination is especially relevant in crowded stellar fields and for instruments with relatively large pixel scales, such as TESS's 21" pixels or PLATO's 15" pixel sampling; even modest angular proximity between stars can significantly distort observed flux ratios, diluting transit depths and altering apparent variability amplitudes \citep[][]{nascimbeni2022plato, mantovan2024gaps}. The Gaia mission \citep[][]{prusti2016gaia} has greatly improved the characterization of these environments through high-precision astrometry, photometry, and stellar density maps across the full sky. Dedicated pipelines address contamination for individual missions, but a scalable, mission-agnostic framework for rapid catalog-level assessment is lacking.\\
Previous works have attempted to address aspects of this issue. The PLATOSim framework \citep[][]{jannsen2024platosim} provided an end-to-end simulation of the PLATO camera system to model noise sources, including blending from background stars, while \cite{nascimbeni2025plato} extended this work to assess field selection criteria and contamination statistics for the mission’s first long-pointing fields. \cite{2024CoSka..54b.142D} analyzed ground-based photometric follow-up strategies for PLATO-detected transits, highlighting the role of contamination correction for confirming planetary signals, and \cite{han2025hundreds} further showed that improved contamination treatment can significantly affect the inferred radii of transiting planets, underscoring that contamination correction matters not only for detecting transits but for characterizing planetary systems once detected.\\
The Transiting Exoplanet Survey Satellite (TESS) illustrates contamination's observational impact directly: its wide-field cameras, optimized for high-cadence, all-sky monitoring, frequently record blended light from multiple stars. \cite{hey2024confronting} and \cite{bluhm2020precise} showed that unaccounted flux contamination in TESS light curves can mimic shallow transits or alter phase-curve modulations, while \cite{leonardi2024taste} emphasized the need for robust source deblending when analyzing mono-transit events. A concrete example is the TESS observations of the active M dwarf AD Leo, whose light curve is contaminated by the bright nearby star $\gamma$ Leo through a CCD bleed trail, requiring customized pixel-level extraction to remove the contaminating flux \citep{stelzer2022great}. Blending can produce astrophysical false positives, such as eclipsing binaries masquerading as transits, or obscure genuine small-planet signals \citep[][]{biazzo2022gaps}; quantifying and mitigating contamination is therefore a prerequisite for reliable astrophysical inference.\\
The Ariel mission, designed for atmospheric spectroscopy of transiting exoplanets, places similar emphasis on photometric stability and contamination control. The Ariel Red Book \citep[][]{tinetti2021ariel} sets stringent requirements on background flux characterization, since even minor contamination from nearby stars can affect transit depth precision and inferred atmospheric properties. \cite{tonucci2025analysing} showed that photometric calibration systems for missions like Ariel must model flux contamination to reach the required parts-per-million accuracy, as exoplanet atmosphere retrieval depends on subtle spectral variations that a small fraction of flux blending may significantly bias.\\
The astrometric precision of Gaia DR3, providing positions, magnitudes, and colors for nearly two billion stars \citep[][]{prusti2016gaia}, allows contamination to be assessed at high spatial detail for any sky region \citep[][]{magrini2023hrmos}; cross-matching Gaia and TESS catalogs, for instance, has revealed systematic flux-blending patterns in high-density Galactic regions \citep[][]{nascimbeni2025plato}. However, catalog sizes of hundreds of millions of sources make naive pairwise spatial calculations computationally prohibitive.\\
This scalability challenge motivates tools capable of estimating contamination efficiently. Large photometric missions must evaluate contamination not only for specific targets but statistically across entire survey fields, to optimize pointing strategies, define exclusion zones, or calibrate photometric apertures \citep[][]{kerins2023rosetz, nascimbeni2025plato}. For example, the PLATO field-selection process explicitly incorporates contamination analysis to prioritize low-density fields \citep[][]{nascimbeni2025plato}. Computational efficiency is therefore not a secondary concern: the neighbor-search problem in multi-million-star catalogs scales quadratically with source count ($O(N^2)$) if not properly optimized, and efficient spatial indexing, parallel computation, and precomputed neighbor databases are what make full-sky contamination estimation feasible.\\
Each mission TESS, PLATO, Ariel, or Mauve \citep[][]{saba2025mauvesim} has distinct PSF shapes, pixel geometries, and magnitude sensitivities that define its effective contamination radius; in PLATO, for example, the large pixel scale and wide photometric apertures make contamination correction a mission-critical step in its pre-launch calibration pipeline \citep[][]{nascimbeni2022plato}.\\
Mauve is a small satellite designed for high-cadence ultraviolet and visible spectrophotometry of stellar activity and magnetic variability, complementing broader-band missions such as TESS and PLATO \citep[][]{saba2025mauvesim, 2026RASTI...5ag018M}. Its compact optical design and wide photometric aperture ($\sim$47") make it particularly sensitive to flux contamination: neighboring stars can contribute to the detected flux degrading photometric precision.\\
While existing approaches provide effective solutions for specific missions and instruments, they are generally optimized for individual pipelines or datasets. A general, catalog-driven framework capable of rapidly evaluating contamination across arbitrary target lists, apertures, and observing geometries remains valuable for mission planning. We therefore developed the Photometric Contamination Analyzer Tool (PHOTO-CAT) as part of Mauve's mission-preparation pipeline, evaluating stellar crowding and flux blending for candidate targets from Gaia DR3 data to flag targets that could compromise photometric accuracy or pointing reliability and to support clean-field selection before observations begin. \\
Catalog-based contamination metrics have been developed for previous wide-field surveys. The TESS science pipeline, for example, provides a crowding metric (CROWDSAP) estimating the fraction of aperture flux from the target star \citep[][]{kourniotis2025variability,calcaferro2024pulsating}. More detailed tools model individual neighboring sources: the \texttt{eleanor} pipeline performs image-level corrections and aperture optimization for TESS full-frame image (FFI) photometry \citep[][]{feinstein2019eleanor}; the TESS-Gaia Light Curve (TGLC) pipeline incorporates Gaia catalogs to model contaminats and extract deblended TESS light curves from FFI observations \citep[][]{han2023tess}; and TESS-cont uses Gaia catalogs together with the instrument pixel response function to estimate the flux contribution of nearby sources within the photometric aperture \citep[][]{castro2024toi}. \\
Unlike these mission-specific pipelines, designed to improve photometric extraction from existing observations, PHOTO-CAT operates directly on source catalogs for pre-observation contamination assessment, target screening, and survey planning. It requires no image data and can therefore be applied before observations are acquired, or to instruments without dedicated photometric pipelines, making it readily transferable to future missions and custom observing configurations.\\
Although developed for Mauve, PHOTO-CAT's modular, catalog-driven architecture is readily adaptable to other wide-field missions such as PLATO, Ariel, or TESS, where similar field-selection and calibration challenges arise. Adapting PHOTO-CAT to a new mission requires specifying the relevant field of view, aperture, and mission-specific contamination thresholds, together with a description of the instrumental response: the PSF characteristics (e.g. its FWHM, derived from the optical design or from calibration data of the instrument) when the Gaussian PSF model is adopted, and the instrumental transmission curve when the photometric band conversion is required. The core computation, based on angular separations and relative brightnesses, is instead universal across instruments.\\
The paper is structured as follows. Section~\ref{sec:workflowstructure} describes PHOTO-CAT's architecture and workflow, from the stellar neighborhood index to contamination-metric computation and output generation. Section~\ref{sec:all-sky} presents an all-sky contamination assessment using Gaia (G$\leq$12, with neighbors down to G=17) and a case study applying PHOTO-CAT to the Ariel Mission Candidate Sample. Section~\ref{sec:Performance_and_Operational_Considerations} discusses performance and scalability, and Sect.~\ref{sec:Tool_Limitations} discusses the main limitations of the current framework and the planned developments. Section~\ref{sec:Conclusions} summarizes the main results.
\section{Workflow structure}\label{sec:workflowstructure}
\begin{figure*}
\centering
\definecolor{pcNavy}{HTML}{1034A6}
\definecolor{pcSky}{HTML}{6FA8DC}
\definecolor{pcGold}{HTML}{E3A92B}
\definecolor{pcRed}{HTML}{C8372D}
\definecolor{pcTeal}{HTML}{1A9E8F} 
\begin{tikzpicture}[
    every node/.append style = {align=center, font=\small,
        execute at begin node = {\hyphenpenalty=10000}},
    main/.style = {text width=6.4cm},
    side/.style = {text width=3.8cm},
    startstop/.append style = {fill=pcRed!55, minimum height=0.8cm},
    process/.append style = {fill=pcSky!65, minimum height=0.8cm},
    decision/.append style = {fill=pcGold!80, minimum height=0.8cm},
    process/.append style = {rounded corners=2pt},
    decision/.append style = {chamfered rectangle, chamfered rectangle xsep=3mm},
    panel/.style = {rectangle, rounded corners=6pt, fill=pcTeal!12, draw=pcTeal!80, semithick, inner sep=3mm},
    ptitle/.style = {font=\small\bfseries, text=pcTeal!75!black}]

\node (start) [startstop, main]
    {START: Load configuration};
\node (logo) [above=3mm of start]
    {\includegraphics[width=3.2cm]{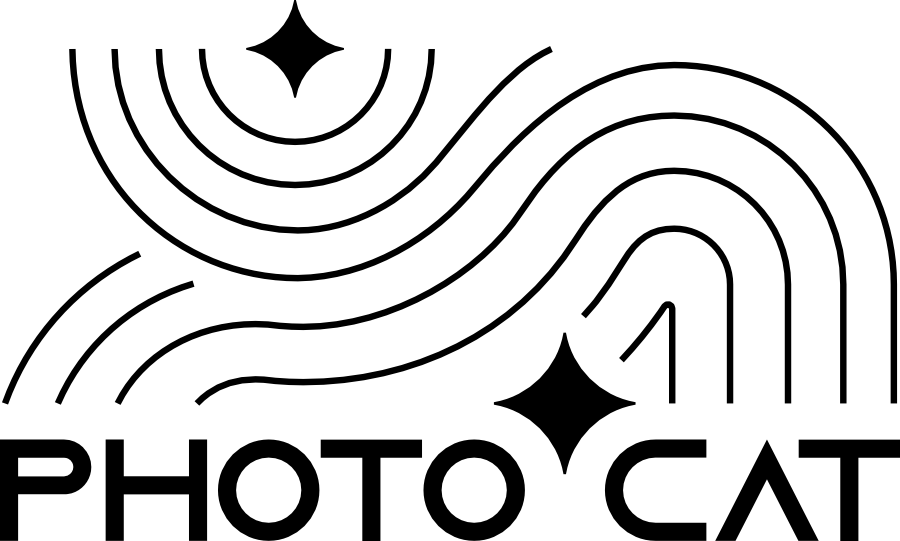}};
\node (idx) [decision, main, below=9mm of start]
    {Index exists for catalog and settings?};
\node (reuse) [process, side, right=9mm of idx]
    {Reuse existing index};
\node (cat) [process, main, below=5mm of idx]
    {Load and clean input catalog};
\node (tree) [process, main, below=5mm of cat]
    {KDTree neighbor search on 3D unit vectors};
\node (csr) [process, main, below=5mm of tree]
    {Save CSR neighbor index and master catalog};

\node (targets) [process, main, below=12mm of csr]
    {Load targets and match source IDs};
\node (conv) [decision, main, below=5mm of targets]
    {Band conversion requested?};
\node (convert) [process, side, right=9mm of conv]
    {Convert magnitudes to output band};

\node (nbr) [process, main, below=6mm of conv]
    {Retrieve target neighbors and separations};
\node (model) [decision, main, below=5mm of nbr]
    {Aperture model?};
\node (tophat) [process, text width=3.4cm, below=8mm of model, xshift = -2.2cm]
    {Top-hat};
\node (psf) [process, text width=3.4cm, below=8mm of model, xshift = 2.2cm]
    {Gaussian PSF};
\coordinate (mid) at (tophat.south -| model.south);
\node (more) [decision, main, below=7mm of mid]
    {More targets?};
\coordinate (loopx) at ($(more.west -| tophat.west)+(-0.7,0)$);
\node (end) [startstop, main, below=10mm of more]
    {END: Save results and summary};

\begin{scope}[on background layer]
\node (p1) [panel, fit=(idx)(reuse)(cat)(tree)(csr)] {};
\node (p2) [panel, fit=(targets)(convert)(nbr)(model)(tophat)(psf)(more)(loopx)] {};
\end{scope}
\node [ptitle, anchor=south west] at (p1.north west) {Index construction};
\node [ptitle, anchor=south west] at (p2.north west) {Per-target processing};

\draw [arrow] (start) -- (idx);
\draw [arrow] (idx) -- node[anchor=south]{yes} (reuse);
\draw [arrow] (idx) -- node[anchor=east]{no} (cat);
\draw [arrow] (cat) -- (tree);
\draw [arrow] (tree) -- (csr);
\draw [arrow] (csr) -- (targets);
\draw [arrow] (reuse) |- (targets);
\draw [arrow] (targets) -- (conv);
\draw [arrow] (conv) -- node[anchor=south]{yes} (convert);
\draw [arrow] (conv) -- node[anchor=east]{no} (nbr);
\draw [arrow] (convert) |- (nbr);
\draw [arrow] (nbr) -- (model);
\draw [arrow] (model.south) -- (tophat.north);
\draw [arrow] (model.south) -- (psf.north);
\draw [arrow] (tophat.south) -- (more.north);
\draw [arrow] (psf.south) -- (more.north);
\draw [arrow, dashed] (more.west) -- node[anchor=south]{yes} (loopx)
    |- node[pos=0.25, sloped, below]{next target} (nbr.west);
\draw [arrow] (more) -- node[anchor=east]{no} (end);

\path let \p1 = (current bounding box.east), \p2 = (current bounding box.west)
    in (-\x1,0) (-\x2,0);
\end{tikzpicture}
\caption{Workflow chart of PHOTO-CAT.}
\label{fig:workflow}
\end{figure*}

This section describes PHOTO-CAT's workflow: construction of the stellar neighborhood index (Sect.~\ref{sec:Building_stellar_neighborhood_index}), target ingestion and matching (Sect.~\ref{sec:Loading_and_matching_targets}), optional photometric band conversion (Sect.~\ref{sec:bandpass_conversion}), computation of contamination metrics under the top-hat and Gaussian PSF aperture models (Sect.~\ref{sec:Computing_contamination_metrics}), and output generation (Sect.~\ref{sec:Saving_and_summarizing_results}).\\
PHOTO-CAT follows a sequential workflow, displayed in Fig.~\ref{fig:workflow}, grouped into four stages:
\begin{enumerate}
    \item Building the stellar neighborhood index: Querying the full catalog to identify, for each catalog star, nearby sources within a fixed angular radius, and storing their astrometric and photometric parameters in an index for efficient access.
    \item Loading and matching targets: Reading the list of target stars, cross-matching their identifiers with the catalog, and preparing the corresponding subset of the neighborhood index for subsequent analysis.
    \item Computing contamination metrics: For each target star, optionally converting the catalog magnitudes to the passband of a selected observing mission before estimating the flux contribution from neighboring sources using either a top-hat aperture model or a circular Gaussian PSF model. The tool computes quantitative contamination metrics, including contaminating flux fractions and, for the Gaussian PSF model, contamination ratio and target purity.
    \item Saving and summarizing results: Storing per-target contamination data in a table.
\end{enumerate}

\subsection{Building the stellar neighborhood index}\label{sec:Building_stellar_neighborhood_index}
The first stage constructs a compact, query-efficient index of the angular environment surrounding each catalog star, enabling rapid retrieval of neighbors within a fixed angular radius. The process combines a Euclidean sky representation, spatial indexing, and streaming I/O to process catalogs containing tens to hundreds of millions of stars without requiring the full dataset to be loaded into memory at once, while also supporting resumable execution.\\
The celestial coordinates of each source, Right Ascension (RA) and Declination (Dec), are mapped onto three-dimensional Cartesian unit vectors on the celestial sphere. This representation provides a Euclidean metric in which angular separations correspond to distances between vectors, avoiding issues related to the periodicity of RA and the non-uniform metric of spherical coordinates. The resulting set of unit vectors is indexed using a k-dimensional tree (KDTree) \citep[][]{bentley1975multidimensional}, reducing the computational cost from naive pairwise comparisons to logarithmic scaling with catalog size. For each source, the algorithm queries the tree to retrieve all neighboring stars within a specified maximum angular radius, storing the resulting neighbor lists and separations in a compressed sparse row (CSR)-like format (Sect.~\ref{sec:CSR-style_storage_and_checkpointing}), allowing downstream tools to reconstruct the local sky environment of any target star without repeating the spatial search.
\subsubsection{Loading and preparing the input catalog}\label{sec:Loading_input_catalog}
\begin{figure}
    \includegraphics[width=\hsize]{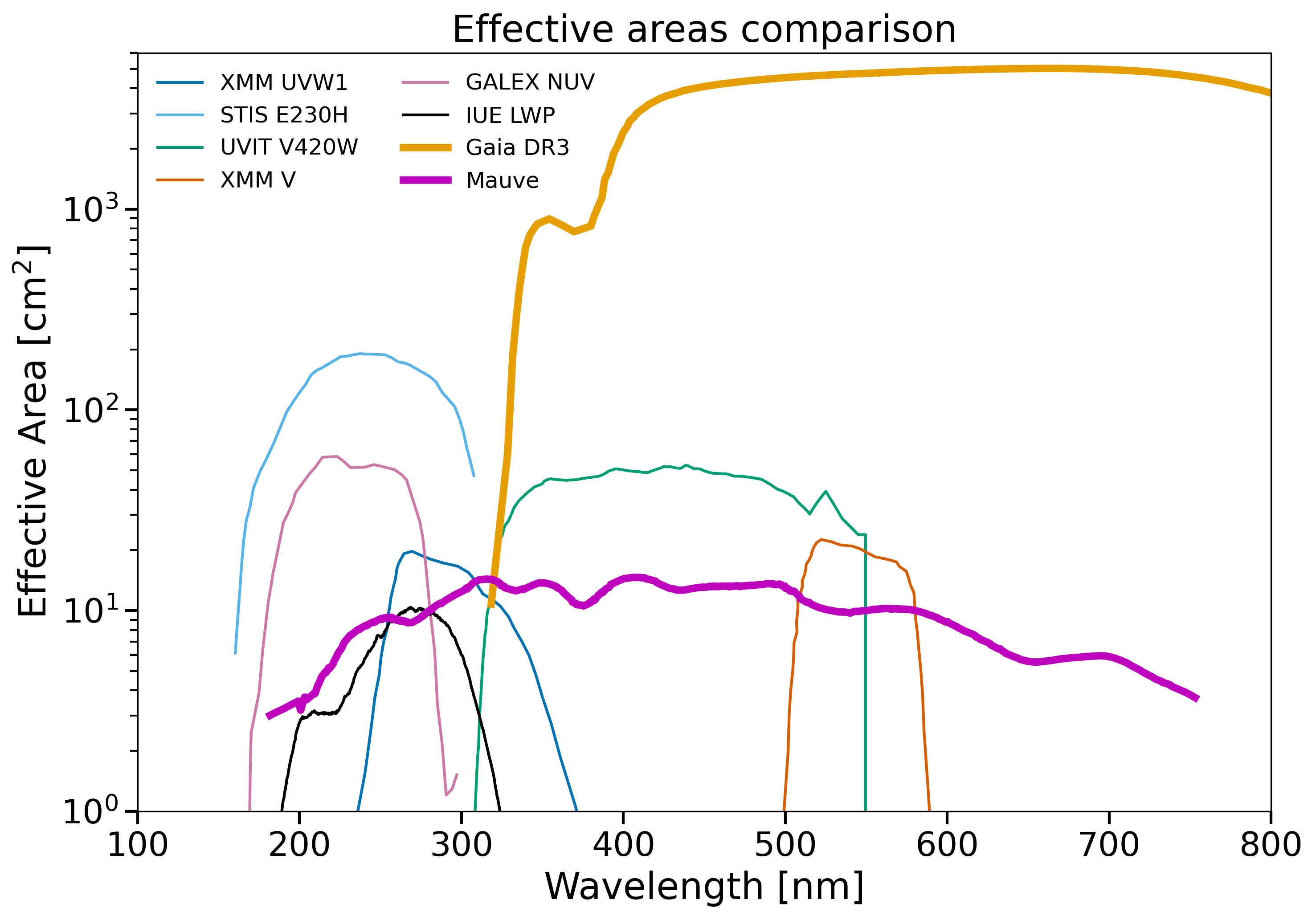}
    \caption{Effective area as a function of wavelength for Mauve (thick magenta line) compared with the Gaia DR3 photometric band (thick orange line). The figure also includes the effective area curves of other representative instruments for reference.}
    \label{fig:effective_area}
\end{figure}

The procedure begins by reading a star catalog, typically extracted from Gaia or a similar astrometric survey. The index requires four columns: the source identifier, RA, Dec, and the apparent magnitude in a reference band such as Gaia magnitude.
In this work, we adopt the Gaia catalog and use the Gaia broad-band magnitude, motivated by the overlap between the Gaia band ($\sim$300--1100 nm) and the Mauve ultraviolet--optical range ($\sim$200--700 nm; Fig.~\ref{fig:effective_area}).
Depending on catalog size, the code uses either Pandas \citep[][]{mckinney-proc-scipy-2010} for in-memory operations, or Dask \citep[][]{dask} for out-of-core, chunked, parallelized processing of catalogs exceeding available memory (tens to hundreds of millions of entries). After loading, the code retains only valid entries (removing missing or invalid coordinates), keeps RA/Dec as floating-point values, and converts source identifiers to 64-bit integers for consistent indexing and efficient binary storage.
\subsubsection{Geometric representation in 3D Euclidean space}\label{sec:Geometric_representation}
Angular separations on the celestial sphere can be computed directly from RA and Dec. However, the wrap-around of right ascension at 360° and the shrinking of RA-distances toward the celestial poles complicate large-scale nearest-neighbor searches. The algorithm therefore maps each star onto a 3D Cartesian unit vector, in which Euclidean distance is directly related to angular separation, enabling fast KDTree-based searches.
For a star at RA = $\alpha$ and Dec = $\delta$, the corresponding unit vector $u = (x, y, z)$ is defined as:
\begin{equation}
    x=\cos{\delta}\cos{\alpha},\,\,\,\, y=\cos{\delta}\sin{\alpha}, \,\,\,\,\, z=\sin{\delta}
\end{equation}
Given two stars with unit vectors $u_1$ and $u_2$, their angular separation $\theta$ is given by:
\begin{equation}
    \cos{\theta}=u_1 \cdot u_2
\end{equation}
\subsubsection{From angular radius to chordal distance}\label{sec:angular_radius_to_chordal}
The neighbor search is performed in Euclidean space using the chordal distance, which is the straight-line distance between two points on the unit sphere, rather than the angular separation along the surface. For two unit vectors $u_1$ and $u_2$, the chordal (Euclidean) distance is
\begin{equation}\begin{split}
    r_{\mathrm{chord}} = \|u_1 - u_2\| 
    =& \sqrt{(x_1-x_2)^2+(y_1-y_2)^2+(z_1-z_2)^2}=\\
    =& 2\sin\left(\frac{\theta}{2}\right),
\end{split} 
\end{equation}
where $\theta$ is the angular separation between the two sources.\\
This transformation allows nearest-neighbor searches to be performed efficiently in Cartesian coordinates, since an angular search radius $\theta_{\max}$ can be expressed as a scalar chordal radius defining the KDTree query region. The corresponding conversion is
\begin{equation}
    r_{\mathrm{chord}}^{\max}
    = 2 \sin\left(\frac{\theta_{\max}}{2}\right),
\end{equation}

This step ensures that the Euclidean-space search is geometrically equivalent to an angular cone search on the sky.
\subsubsection{Building or loading the KDTree}\label{sec:kdtree}
Once all stars are represented by their 3D coordinates, the code constructs a KDTree using the cKDTree implementation available in SciPy \citep[][]{virtanen2020scipy}, which provides a highly optimized backend for large-scale nearest-neighbor queries. The KDTree recursively partitions the space along coordinate axes, allowing logarithmic-time neighbor searches. Specifically, for a query point $q$, the KDTree identifies all points $p_i$ within a radius $r$ by traversing only a fraction of the tree nodes, rather than comparing $q$ to every catalog entry. This reduces the computational cost of neighborhood searches from $O(N^2)$ to approximately $O(N \log N)$ for tree construction and $O(\log N)$ per query on average.\\
The code saves the tree to disk after initial construction. If the tree already exists, it is automatically reloaded.
\subsubsection{Streaming neighbor search and memory optimization}\label{sec:Streaming_neighbor_search}
The most critical and computationally intensive part of the process is the neighborhood search itself. For each star in the catalog, the KDTree is queried to find all sources within the pre-defined chordal distance. However, since the number of stars can easily reach tens of millions, storing all neighbors in memory simultaneously is not convenient.\\
To solve this, the algorithm operates in streaming mode. The catalog is divided into chunks of configurable size (e.g. 10,000 or 50,000 rows), and each chunk is processed sequentially. For every block of stars, the corresponding 3D coordinates are passed to the tree’s $query\_ball\_point$ method, which returns lists of indices representing the neighbors of each star within the given radius.\\
Each neighbor list is immediately translated from internal indices to physical source IDs using the catalog’s source id column. These IDs are then appended to a temporary buffer. To avoid excessive memory usage, the buffer is periodically flushed to disk in binary format. The same logic applies to optional separation values, which are stored if requested. This streaming design allows the code to scale linearly with catalog size, with minimal peak memory footprint.
\subsubsection{Computing angular separations}\label{sec:Computing_angular_separations}
When the user requests the computation of angular separations, the code efficiently derives them from the unit vectors without resorting to slow trigonometric conversions for every pair. For a given target star with unit vector $u_t$ and its set of neighbors $u_n$, the separations $\theta$ are computed as:
\begin{equation}
    \theta= \arccos{u_t \cdot u_n}
\end{equation}
The dot products are calculated using vectorized operations.\\
This supports downstream contamination models that depend on both flux and distance weighting.

\subsubsection{CSR-style storage and checkpointing}\label{sec:CSR-style_storage_and_checkpointing}
To efficiently encode the variable-length neighbor lists, the algorithm adopts a CSR–like format, a data structure commonly used for sparse matrices. In this layout:
\begin{itemize}
    \item The source IDs array stores the list of all catalog stars, ordered as in the input catalog.
    \item The offsets array (of length $N + 1$) acts as an index table: each entry $\mathrm{offsets}[i]$ stores the position in the binary neighbor file where the neighbor list of the $i$-th star begins. The last element, $\mathrm{offsets}[N]$, corresponds to the total number of neighbors stored. The difference $\mathrm{offsets}[i+1] - \mathrm{offsets}[i]$ therefore gives the number of neighbors associated with star $i$.
    \item The binary neighbor file concatenates all neighbor IDs sequentially as 64-bit integers, without delimiters or padding.
\end{itemize}
This structure allows each star’s neighbors to be retrieved in constant time, without scanning the entire dataset, and minimizes storage overhead compared to explicit matrix formats.

To guarantee robustness against interruptions, the code periodically saves a checkpoint file that records the index of the last processed star, the current offsets array, and the total number of neighbors written so far. If execution is interrupted (for example due to system limits or user termination), the process can resume from the last saved state without data loss or reprocessing of completed stars.
\subsubsection{Compact ID Mapping}\label{sec:Compact_ID_Mapping}
Large astronomical catalogs frequently contain heterogeneous identifiers: numerical IDs (e.g. Gaia DR3 source id) alongside alphanumeric names (e.g. HD 216608A, TYC 1234-567-1). Storing such IDs as strings is costly at scale (memory fragmentation, slower indexing and serialization); PHOTO-CAT therefore introduces a compact internal identifier,
\begin{equation}
    \mathrm{internal\_id}\in\{1,2,....,N\}
\end{equation}
where $N$ is the total number of stars in the input catalog.
Each catalog row is assigned a unique internal id, and every internal structure, arrays, CSR offsets, binary neighbor lists, and KDTree indices, is built exclusively on these integer IDs.\\
Using internal id also enables constant-time access to all geometric and photometric quantities. Because each internal identifier maps directly to a row index ($index=internal\_id-1$), the tool can retrieve RA, Dec, magnitudes, or any additional stored quantity without invoking dictionary lookups or string comparisons. This direct mapping ensures optimal use of NumPy’s vectorized operations, which rely on simple pointer arithmetic and benefit from cache-friendly memory layouts. The CSR neighbor index naturally aligns with this scheme: every neighbor list is expressed in terms of internal id, allowing the algorithm to traverse the binary storage in a strictly sequential fashion and minimizing disk I/O overhead when streaming billions of bytes from the neighbor file.\\
To maintain compatibility with external catalogs, the tool separates internal from external identifiers. Numeric IDs are stored in a compact int64 array, indexed by the corresponding internal id, enabling a trivial and extremely fast mapping back to the original catalog. A small minority of objects, whose identifiers are non-numeric (for example “HD 216608A”), are handled separately in a dedicated structure that stores only the relevant internal id values and their original string names. This hybrid mapping preserves full fidelity to the external catalog while ensuring that only a tiny fraction of string data is ever loaded into memory. As a result, the tool retains the flexibility to work with heterogeneous datasets without compromising the performance advantages of an integer-based internal representation.

\subsubsection{Output summary and master catalog}\label{sec:Output_summary_and_master_catalog}
After completing the neighbor searches, the temporary binary files are automatically renamed to their final versions, ensuring that partial results are never mistaken for complete outputs. The master catalog, containing only the essential columns required for subsequent processing, is saved in Parquet format, a columnar storage format designed for efficient large-scale data analysis. The file is compressed using the Snappy compression algorithm, a fast, lightweight, and lossless compression method optimized for high throughput rather than maximum compression ratio. This combination ensures compact storage while maintaining rapid read/write performance for downstream applications. Finally, the script logs the total number of stars, the number of neighbor entries written, and the output paths of all generated files.
\subsection{Loading and matching targets}\label{sec:Loading_and_matching_targets}
Once the stellar neighborhood index has been built, the contamination tool loads the global catalog and associates each target star with its corresponding entry in that index. 
Targets can be provided either as a CSV file containing a source id column or as a Python list of identifiers passed directly through the configuration. The loader validates that each requested source ID exists in the catalog; missing entries are recorded as "missing" with null astrometric/photometric fields, preserving the one-to-one mapping between requested and processed targets.\\
Decoupling the target list from the index improves modularity: the same index serves multiple target lists sharing the same parent catalog, and supports distributed processing, since worker processes can independently access catalog entries from shared memory via the id-to-index mapping.

\subsection{Photometric band conversion}\label{sec:bandpass_conversion}
The contamination metrics described in Sect.~\ref{sec:Computing_contamination_metrics} are, by default, computed directly from the input catalog's magnitudes (e.g. Gaia $G$). Because stellar contamination is wavelength dependent. Stars with different spectral energy distributions (SEDs) contribute different relative fluxes in different instrumental bandpasses. Two neighboring stars with identical $G$ magnitudes but different colors can produce markedly different contamination levels in the ultraviolet, optical, or infrared.\\
PHOTO-CAT can therefore optionally convert catalog magnitudes into the target instrumental band before evaluating contamination, leaving the algorithms of Sect.~\ref{Sec:tophat} and Sect.~\ref{sec:psfmodel} unchanged: only the flux assigned to each source is modified. The conversion is based on three successive steps:
\begin{enumerate}
    \item stellar color is computed from the available catalog photometry (e.g. BP$-$RP for Gaia DR3).
    \item the color is converted into an equivalent stellar temperature by assuming a parameterized spectral energy distribution.
    \item the resulting spectrum is integrated through both the catalog and the requested instrumental passbands to determine the corresponding flux ratio.
\end{enumerate}
The conversion depends on the adopted description of the source spectrum. We have implemented one empirical, non-SED approach and one SED-based approaches:
\begin{enumerate}
    \item Empirical Gaia photometric relations (Sect. \ref{sec:gaia_empirical})
    \item an atmospheric model (Sect. \ref{sec:atmosphere_conversion})
\end{enumerate}
Because PHOTO-CAT computes contamination from relative flux ratios, only differential magnitudes enter the calculation, so the absolute photometric zero point cancels and the conversion depends only on the relative flux scaling between passbands.\\
Instrumental transmission curves are tabulated wavelength--throughput pairs; PHOTO-CAT includes curves for Gaia, TESS, CHEOPS, Ariel, and Mauve, and accepts user-supplied filters without modification of the underlying algorithm. Passband integrals are precomputed over a temperature grid, so each source requires only interpolation between the precomputed values, preserving the tool's linear scaling with catalog size.
\subsubsection{Empirical Gaia photometric relations}\label{sec:gaia_empirical}
As an alternative to an explicit SED model, PHOTO-CAT implements the empirical photometric relationships provided in the Gaia DR3 documentation. These relations were derived from Gaia DR3 cross-matches with several established photometric systems, including Hipparcos, Tycho-2, SDSS12, Johnson-Cousins, 2MASS, and GSC2.3. The transformations are expressed as polynomial functions of Gaia or external photometric colors and provide direct estimates of photometry in the corresponding target system. \citep{riello2021gaia,gaiaDR3photometricrelations}\\
The empirical relation are expressed in the general form:
\begin{equation}
G-X =
\sum_{i=0}^{N} a_i C^i,
\end{equation}
where $C$ is the calibrated color and $X$ is the magnitude in the target photometric system.\\
The coefficients $a_i$ and the associated color intervals are tabulated per various colors interval in the published Gaia DR3 relations. The Gaia documentation explicitly defines validity ranges for each transformation; the empirical relation is therefore applied only within its calibrated color interval and is not extrapolated outside that range. The original calibrations were constructed using carefully selected Gaia DR3 sources with high-quality photometry and additional filtering criteria designed to reduce photometric noise and excess-flux contamination. \citep{riello2021gaia,gaiaDR3photometricrelations}\\
Unlike the SED-based method, the empirical approach does not infer an effective temperature and does not require an assumed stellar spectral model. Instead, it directly maps the available photometric color to the requested magnitude through the published calibration. This makes the method computationally inexpensive and avoids the model dependence associated with atmosphere-based SEDs.\\
The empirical method is necessarily limited to photometric systems for which a Gaia calibration is available. It is therefore complementary to the SED-based conversion: when a published Gaia relation exists and the source color lies within its calibrated range, PHOTO-CAT can use the empirical transformation directly; for arbitrary instrumental passbands, such as mission-specific filters without a corresponding published Gaia relation, the SED-based method remain the applicable approach. The latter can use an arbitrary transmission curve because the conversion is performed through direct integration of the adopted SED over the input and output passbands.\\
Because the Gaia transformations are empirical photometric relations rather than physical SED models, the resulting magnitudes should be interpreted as calibrated photometric estimates rather than predictions of the underlying stellar spectrum. In particular, the Gaia documentation notes that different stellar populations can exhibit different loci in color--color space and that some extensions to very red colors are not equally representative of dwarfs and giants.\citep{riello2021gaia,gaiaDR3photometricrelations}

\subsubsection{Model-atmosphere spectral conversion}\label{sec:atmosphere_conversion}
PHOTO-CAT can perform photometric band conversion using synthetic spectra from stellar-atmosphere model grids. In this approach, the stellar spectral energy distribution is represented by a model atmosphere characterized by the stellar effective temperature, surface gravity, and metallicity. This provides a more physically detailed wavelength-dependent description of the stellar flux.\\
PHOTO-CAT combines three atmosphere-model families to cover the range
$2000 \leq T_{\rm eff} \leq 55000$ K:
\begin{enumerate}
\item BT-Settl models for $T_{\rm eff}<7000$ K;
\item ATLAS9 models for $7000\leq T_{\rm eff}<15000$ K;
\item TLUSTY models for $15000\leq T_{\rm eff}\leq55000$ K, combining the BSTAR2006 and OSTAR2002 grids.
\end{enumerate}
The BT-Settl models provide synthetic spectra for cool stars and include the treatment of dust formation and settling in the atmosphere \citep{allard2012models}. The intermediate-temperature regime is covered by the ATLAS9 atmosphere grid of \citet{2003IAUS..210P.A20C}. At higher temperatures, PHOTO-CAT uses the non-local thermodynamic equilibrium (non-LTE), line-blanketed TLUSTY models, which account for the effects of numerous atomic absorption lines on the emergent stellar spectrum, combining the BSTAR2006 and OSTAR2002 grids \citep{lanz2003grid,lanz2007grid}. These grids provide complementary coverage of hot-star atmospheres and are accessed through the corresponding Spanish Virtual Observatory (SVO) theoretical spectra collections \citep[][]{rodrigo2020svo}.\\
To assess whether the combination of different atmosphere-model families could introduce artificial discontinuities in the resulting broad-band fluxes, we also tested the models at $T_{\rm eff}=7000$ K and $15000$ K. For a fixed representative set of $\log g=4.5$ and $[{\rm M/H}]=0$, the flux predicted by the two adjacent atmosphere families was integrated directly through the Ariel transmission curves. The relative differences between the models at the nominal transition temperatures were $0.36\%$ at 7000 K and $2.97\%$ at 15000 K for the VISPhot band (the Ariel visible photometric channel), and $0.31\%$ and $1.88\%$, respectively, for AIRS-CH0 (the 1.95--3.9 $\mu$m channel of the Ariel InfraRed Spectrometer). These differences indicate that the model transitions introduce only small broad-band offsets in the tested Ariel passbands, with the largest mismatch remaining below $3\%$. This test is specifically intended to diagnose the consistency of the model joins at the temperature boundaries; it is not a validation of the individual atmosphere models themselves, and is restricted to the two Ariel passbands and the representative set of atmospheric parameters considered here.\\
For each source, the atmosphere model is selected from the stellar parameters $T_{\rm eff},\,\log g,\,[{\rm M/H}]$.\\
The preferred atmosphere family is first determined from $T_{\rm eff}$. PHOTO-CAT then verifies that the requested combination of $T_{\rm eff}$, $\log g$, and $[{\rm M/H}]$ is covered by the selected grid. This explicit coverage check is required because atmosphere grids are discrete and may contain gaps or boundaries in parameter space.\\
When the requested parameters fall between available grid nodes, PHOTO-CAT performs multidimensional interpolation between the neighboring atmosphere models. The interpolated spectrum can be expressed schematically as
\begin{equation}
s_{\lambda}=\mathcal{I}\left(T_{\rm eff},\log g,[{\rm M/H}]\right),
\end{equation}
where $\mathcal{I}$ denotes the interpolation over the surrounding grid nodes. Interpolation is performed only when the complete set of required neighboring nodes is available, preventing the procedure from crossing gaps in the model grid or combining disconnected regions of parameter space.\\
The photometric signal in a passband $b$ is then obtained by integrating the selected or interpolated atmosphere spectrum through the corresponding transmission curve,
\begin{equation}
S_b =\int s_{\lambda}T_b(\lambda)\mathrm{d}\lambda .
\end{equation}
where the SED is expressed in photon-rate units appropriate for photon-counting detectors.
The ratio between the flux transmitted through the requested observing band ("out") and the catalog band ("in") is

\begin{equation}
R=
\frac{S_{\rm out}}
{S_{\rm in}}.
\end{equation}
The magnitude in the requested band is then obtained as

\begin{equation}
m_{\rm out}
=
m_{\rm in}
-
2.5\log_{10}R.
\end{equation}
The atmosphere-model conversion is more computationally demanding than the empirical relations because the required synthetic spectra have to be retrieved and interpolated from the atmosphere grids. Once the relevant spectra have been obtained, the synthetic photometry is evaluated independently for each source and for each requested transmission curve.\\
Unlike the empirical relations described in Sect.~\ref{sec:gaia_empirical} the atmosphere-based method can be applied to arbitrary user-supplied transmission curves through direct integration of the synthetic stellar spectrum.
\subsection{Computing contamination metrics}\label{sec:Computing_contamination_metrics}
The core of the algorithm lies in the computation of contamination metrics for each target star, combining catalog-based photometry, angular separations, and a configurable aperture model. Two aperture models are implemented: a top-hat model, where all sources within a user-defined aperture contribute their full flux, and a circular Gaussian PSF model, which radially weights contaminating sources via a simplified PSF description parameterized by its full width at half maximum (FWHM).
For each target, the code retrieves the contiguous block of neighbor identifiers stored in the CSR-like binary structure built previously (Sect.~\ref{sec:CSR-style_storage_and_checkpointing}):

\begin{equation}
    \mathrm{neighbors}_i = \mathrm{neighbors}_{ids}[\mathrm{offsets}[i] : \mathrm{offsets}[i+1]]
\end{equation}
This array is accessed through NumPy memory mapping ($np.memmap$) \citep[][]{harris2020array}, loading data lazily from disk to enable efficient processing of catalogs with billions of neighbor links without excessive RAM usage.
For each contaminant candidate, the angular separation from the target is computed from their Cartesian unit vectors on the celestial sphere,
\begin{equation}
\theta =
\arccos\left(\hat{\mathbf r}_{\rm target}\cdot
\hat{\mathbf r}_{\rm cont}\right)
\end{equation}
where each unit vector is obtained from the corresponding right ascension and declination.
This avoids small-angle-approximation errors near the celestial poles. The resulting separations are filtered to a circular search region defined in the configuration, and the photometric aperture is applied to convert the magnitudes of these sources into relative fluxes.
\subsubsection{Top-hat model}\label{Sec:tophat}
In the top-hat approximation, every source within the aperture radius contributes its full flux regardless of angular distance, a computationally efficient, conservative upper-bound estimate. The contaminating flux fraction is computed as:
\begin{equation}
f_{\mathrm{extra}} = \frac{\sum_i F_i}{F_{\mathrm{target}}}
\end{equation}
where:
\begin{itemize}
\item $F_i = 10^{-0.4m_i}$ is the flux of the $i$-th contaminant
\item $F_{\mathrm{target}} = 10^{-0.4m_{\mathrm{target}}}$ is the flux of the target star
\end{itemize}

This quantity, expressed as a percentage, represents the fractional contribution of all external light sources within the search region to the target’s flux.
We distinguish flux contamination from contaminants throughout this work: flux contamination sums the contribution of every neighboring source in the search region regardless of brightness, whereas the reported number of contaminants counts only sources satisfying the adopted magnitude threshold. A target can therefore show non-zero flux contamination with zero counted contaminants, since sub-threshold neighbors still contribute a small amount of flux.\\
The total $f_{\mathrm{extra}}$ is computed considering all neighboring stars in the FoV, even those exceeding the adopted $\Delta m$ threshold, ensuring that the overall contamination is not underestimated. The algorithm retains only the contaminants that meet a user-defined magnitude contrast criterion:
\begin{equation}
\Delta m = m_{\mathrm{neighbor}} - m_{\mathrm{target}} \leq \Delta m_{\mathrm{max}}
\label{eq:deltam}
\end{equation}
where $\Delta m_{\mathrm{max}}$ is typically set to a few magnitudes (e.g. 2 mag), isolating contaminants bright enough to significantly bias the photometric signal.\\
\subsubsection{Circular Gaussian PSF model}\label{sec:psfmodel}
Unlike the top-hat model, where every source in the aperture contributes its total flux regardless of position, the PSF-based approach weights each contaminant's contribution by the fraction of its flux that falls inside the aperture, which depends on its distance from the aperture center. Each source is represented as a circular 2D Gaussian PSF parameterized by a user-defined FWHM, which must be derived from the optical characteristics of the instrument (e.g. its optical design or measured PSF). The user also sets the maximum PSF extent to consider, expressed as a multiple $N_\sigma$ of the Gaussian standard deviation, converted internally as:
\begin{equation}
    \sigma=\frac{\mathrm{FWHM}}{2\sqrt{2\ln2}},
\end{equation}
giving an effective influence radius
\begin{equation}
    R_{\rm inf}=R+N_\sigma\,\sigma,
\end{equation}
where $R$ is the aperture radius. A source centered outside the aperture still contributes to the collected flux as long as its PSF wings reach the aperture edge, i.e. for distances $d<R+N_\sigma\sigma$ from the aperture center. Neighboring sources within $R_{\rm inf}$, both inside and outside the aperture, are therefore included in the contamination calculation, and the neighborhood search in PSF mode uses $R_{\rm inf}$ rather than the aperture radius alone. Only the fraction of their flux that falls inside the aperture is counted (Eq.~\ref{eq:eta}), so that the out-of-aperture sources contribute the flux spilled into the aperture by their PSF wings. Since the Gaussian profile decreases rapidly beyond a few standard deviations, restricting the search to $R_{\rm inf}$ (typically $N_\sigma\simeq4$--6, beyond which the fraction of the PSF flux of a source that can still reach the aperture is $\lesssim10^{-4}$) preserves accuracy while reducing computational cost for large catalogs.
A source with total flux $F$ and centroid position $\mathbf{r}_0$ is described by the surface brightness distribution:
\begin{equation}
    I(\mathbf{r})=
    \frac{F}{2\pi\sigma^2}
    e^{-\frac{|\mathbf{r}-\mathbf{r}_0|^2}{2\sigma^2}}
\end{equation}
The fraction of each contaminant's flux collected within the aperture radius $R$ depends on its distance $d$ from the aperture center, described by the encircled-energy function:
\begin{equation}
    \eta(d;R,\sigma)=
    F_{\chi^{'2}_2((d/\sigma)^2)}
    \left((R/\sigma)^2\right)
    \label{eq:eta}
\end{equation}

where $F_{\chi^{'2}}$ is the cumulative distribution function of a non-central $\chi^2$ distribution with two degrees of freedom (the non-centrality parameter accounting for the source-aperture offset). For an on-axis source ($d=0$), this reduces to the classical Gaussian encircled-energy fraction:

\begin{equation}
    \eta(0;R,\sigma)
    =
    1-e^{
    -\frac{R^2}{2\sigma^2}
    }
\end{equation}

The relative flux contribution with respect to the target follows from the Pogson relation:

\begin{equation}
\rho_i=
10^{-0.4(m_i-m_t)}
\end{equation}

so the total PSF-weighted contaminating flux, in units of target flux, is:

\begin{equation}
    F_{\rm cont}=
    \sum_i
    \eta(d_i;R,\sigma)
    10^{-0.4(m_i-m_t)}
\end{equation}
where the sum runs over all neighboring sources within $R_{\rm inf}$ satisfying the adopted magnitude contrast criterion (Eq.~\ref{eq:deltam}), irrespective of whether they are centered inside or outside the aperture.

Assuming the target is centered on the aperture, its effective collected flux is the on-axis encircled-energy fraction,

\begin{equation}
    F_{\rm target}=
    1-e^{
    -\frac{R^2}{2\sigma^2}
    }
\end{equation}

PHOTO-CAT reports the contamination ratio

\begin{equation}
C=\frac{F_{\rm cont}}{F_{\rm target}}
\end{equation}
where (C) is the contamination ratio, expressed as a fraction of the target flux. The corresponding target purity is defined as

\begin{equation}
P=
\frac{F_{\rm target}}
{F_{\rm target}+F_{\rm cont}}
=
\frac{1}{1+C}
\end{equation}

where (P) is the fraction of the total collected flux contributed by the target.
The PSF-weighted contaminating flux fractions reported in the output adopt the same weighting, $\eta(d_i;R,\sigma)/\eta(0;R,\sigma)$, for each neighbor, so that they express the flux collected from the contaminants relative to the flux collected from the target and are consistent with $C$.
Each contaminant is stored as a structured record (identifier, coordinates, magnitude, separation); the algorithm also logs the total contaminant count and minimum separation found.\\
This step is computationally efficient: vectorized NumPy operations, pre-indexed lookups, and on-disk memory mapping avoid redundant reads, and the design would allow straightforward GPU acceleration or multiprocessing for larger surveys.

\subsubsection{Visualization of contamination scenarios}
\begin{figure*}
    \centering
    \subfloat[]{
        \includegraphics[width=0.25\hsize]{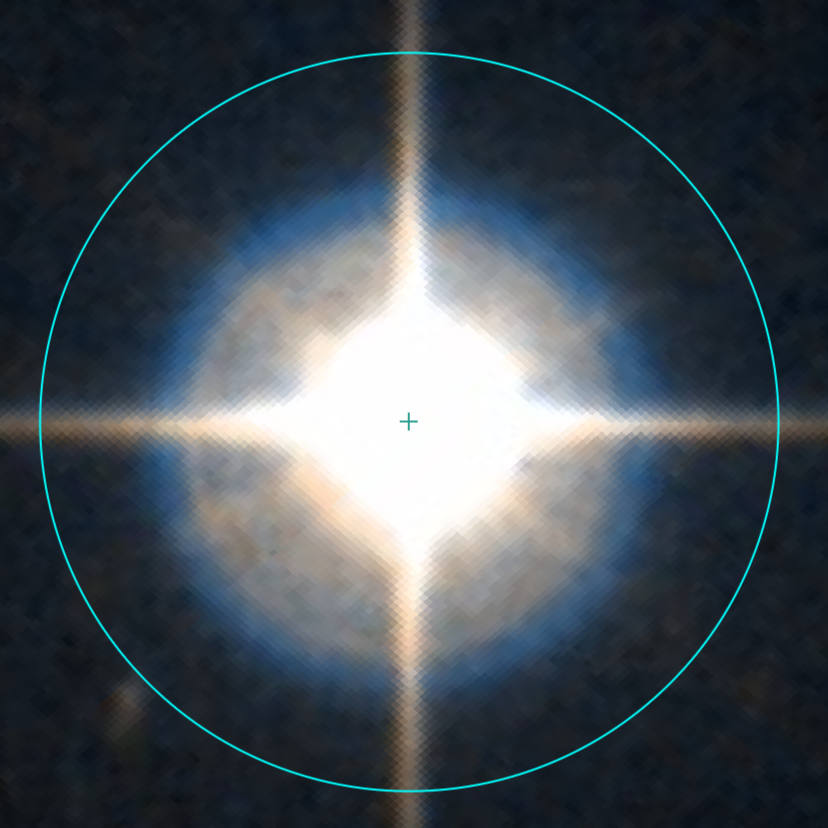}
        \label{fig:2519569450970641408_0_cont_0}
    }
    \subfloat[]{
        \includegraphics[width=0.25\hsize]{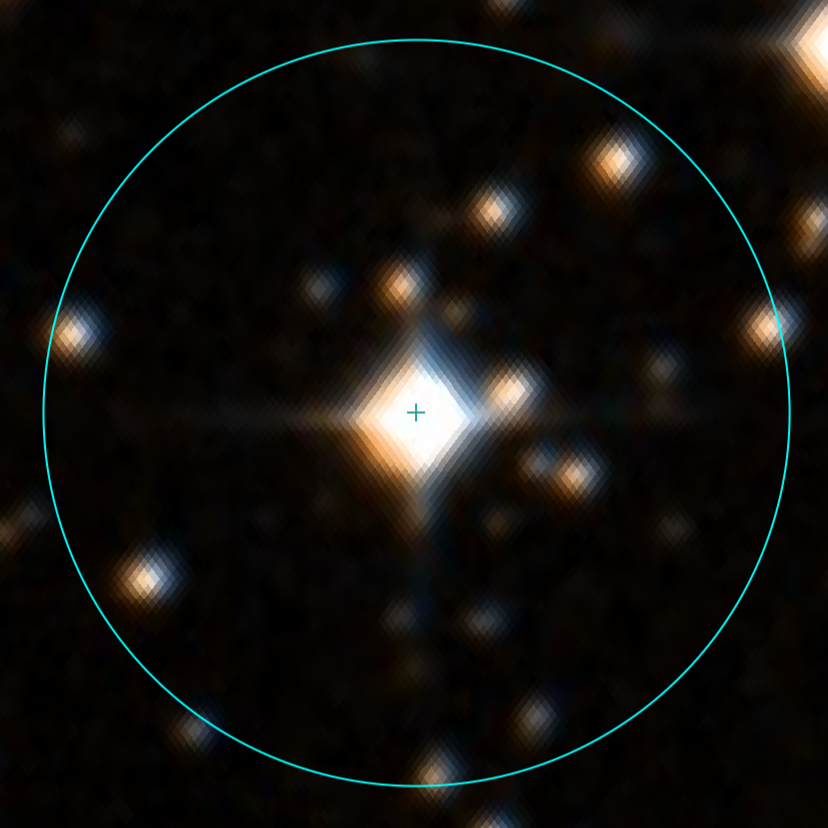}
        \label{fig:/5436012086462390656_no_contaminants_3,4}
    }
    \subfloat[]{
        \includegraphics[width=0.25\hsize]{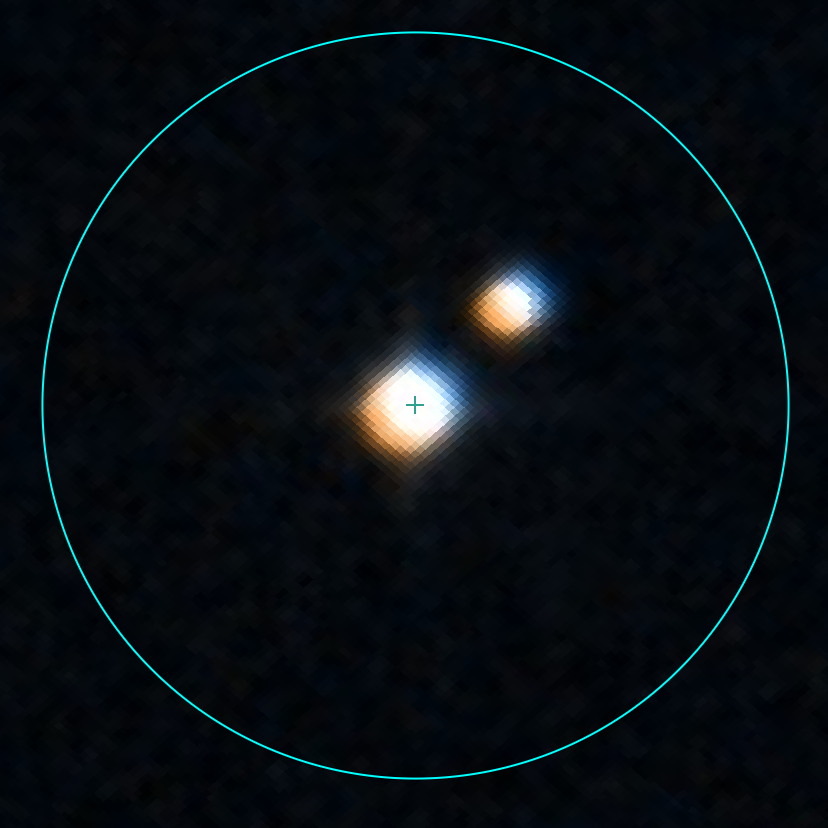}
        \label{fig:2567695379860654848_1_cont_12,78}
    }
    \subfloat[]{
        \includegraphics[width=0.25\hsize]{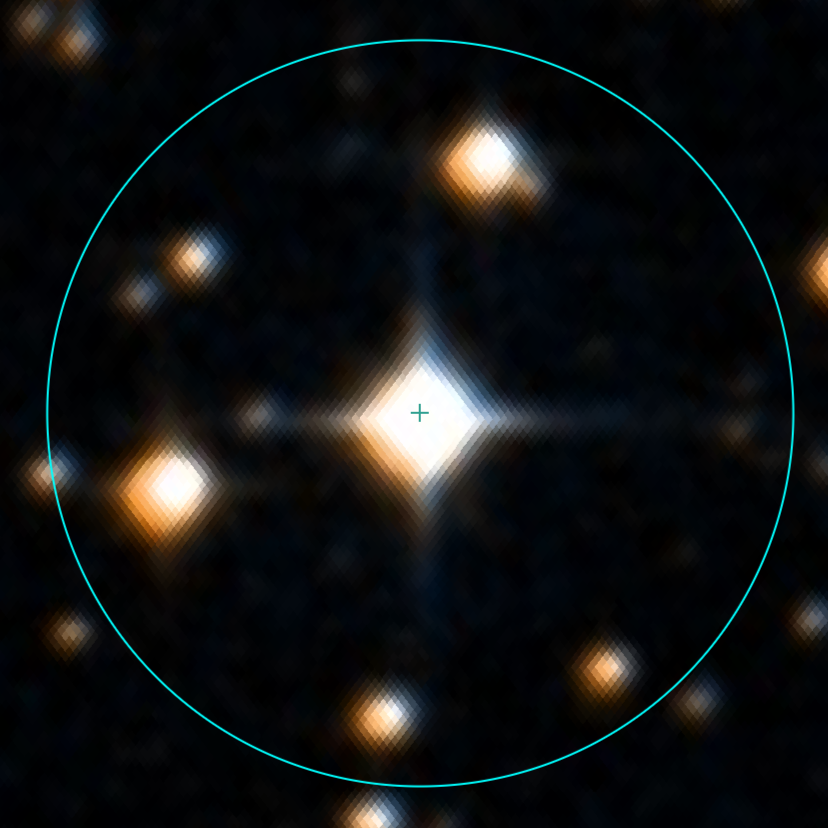}
        \label{fig:5436118636010447360_3_cont_28,13}
    }
    \caption{Examples of contamination scenarios based on DSS images retrieved through the SkyView service. The images cover a $50"\times50"$ field, and the light-blue circle indicates the 47" circular photometric aperture adopted in the top-hat aperture model. The reported flux contamination includes the contribution from all neighboring stars within the aperture, whereas the number of contaminants refers only to neighbors satisfying the adopted magnitude threshold ($\Delta G = 5$). (a) Gaia DR3 2519569450970641408, 0 contaminants, 0\% flux contamination;(b) Gaia DR3 5436012086462390656, 0 contaminants, 3.4\% flux contamination;(c) Gaia DR3 2567695379860654848, one contaminant, 12.78\% flux contamination,(d) Gaia DR3 5436118636010447360, 3 contaminants, 28.13\% flux contamination.}

    \label{fig:contamination_examples}
\end{figure*}

To provide a visual representation of the contamination scenarios, we retrieved archival images from the Digitized Sky Survey (DSS) \citep{lasker1990guide} through the NASA SkyView service \citep{mcglynn1998skyview}. The images shown in Fig. \ref{fig:contamination_examples} are real sky observations, not simulations, displayed as RGB composites generated from DSS2 Red and DSS2 Blue photographic plates (or DSS/DSS2 Blue when RGB data were unavailable). Each image covers a $50"\times50"$ field around the target, with north up and east to the left. The images were retrieved with a pixel scale of approximately 0.2"$ pixel^{-1}$ (600 pixels over a 120" field of view). The light-blue circle represents the 47" circular photometric aperture adopted in the top-hat aperture model.\\
Figure \ref{fig:contamination_examples} presents four representative examples of contamination scenarios within a 47" circular photometric aperture, corresponding to the aperture adopted for the Mauve contamination analysis. The examples illustrate the distinction between the total contaminating flux and the contaminant count.
Fig. \ref{fig:2519569450970641408_0_cont_0}, Gaia DR3 2519569450970641408, illustrates a case with no reported contaminants according to the adopted magnitude-contrast criterion, yet it exhibits a flux contamination of 3.4\%. This contamination arises from neighboring stars that are fainter than the adopted threshold and therefore are not counted as contaminants, although their combined flux still contributes to the total contamination.
Fig. \ref{fig:/5436012086462390656_no_contaminants_3,4}, Gaia DR3 5436012086462390656, shows a star with no detected contaminants within the search region and a contaminating flux fraction of 3.4\%.
Fig. \ref{fig:2567695379860654848_1_cont_12,78}, Gaia DR3 2567695379860654848 exhibits one contaminant producing a 12.78\% flux contribution.
The last example, Fig. \ref{fig:5436118636010447360_3_cont_28,13}, Gaia DR3 5436118636010447360, has three contaminant sources contributing a total of 28.13\% to the observed flux.\\

\subsection{Saving and summarizing results}\label{sec:Saving_and_summarizing_results}
Once contamination values have been computed for all targets, results are serialized as a JSON summary containing, for each target: source identifier, coordinates, catalog magnitude, the photometric band in which contamination was evaluated, the adopted aperture model, the resulting contamination quantities (top-hat flux fraction, or contamination ratio/target purity/PSF-weighted fraction for the Gaussian model), and neighbor separations. When band conversion is enabled, reported magnitudes and metrics refer to the output band rather than the catalog band. For each target, the output also stores the identified neighboring sources and their associated properties, including source identifiers, coordinates, magnitudes, angular separations, aperture classification, and available photometry, allowing individual contamination sources to be identified and the sensitivity of the results to the adopted observing geometry to be assessed. The output filename encodes the run parameters (FoV radius, magnitude threshold, timestamp) for reproducibility.\\
The code also logs post-run summary statistics: total targets processed, count with nonzero contamination, and count without contaminants within the FoV, giving an immediate indication of the field's density regime (e.g. a high contaminated fraction typically signals proximity to the Galactic plane).\\
The modular output structure facilitates integration with external tools, reflecting the tool's design philosophy of transparency and reproducibility: each step, from geometric filtering to flux computation, is explicitly recorded so results can be regenerated under modified assumptions (e.g., different FoV, $\Delta mag$, PSF parameters, or photometric band).
\section{All-sky analysis}\label{sec:all-sky}
This section presents an all-sky contamination assessment using Gaia and supplementary catalogs (sample construction below; global statistics in Sect.~\ref{sec:global_contamination}; contaminant and separation distributions in Sect.~\ref{sec:Distribution_of_contaminating_sources}; spatial sky-map in Sect.~\ref{sec:sky-map}), followed by a case study on the Ariel Mission Candidate Sample (Sect.~\ref{sec:ariel_application}).\\
To evaluate the expected level of stellar contamination in wide-field photometric observations, we applied PHOTO-CAT to an all-sky stellar sample built from the Gaia catalog, quantifying how frequently nearby stars introduce additional flux within a typical photometric aperture and how this depends on stellar magnitude and angular separation. We focus on stars with G $\leq$ 12, roughly matching the brightness range targeted by space-based small-aperture spectrophotometric instruments like Mauve.\\
The target list comprises all-sky stars with $G<12$ (3.088.117 targets). Since nearby faint companions can still contribute significant flux, we queried the Gaia catalog down to $G=17$ ($\Delta G=5$, i.e. 1\% of the target flux), ensuring that all neighbors capable of contributing non-negligible contamination are included. Because Gaia is incomplete for very bright stars ($G\lesssim6$) due to saturation, the catalog was supplemented with the Bright Star Catalog for these sources, giving a homogeneous, near-complete stellar sample over the relevant magnitude range.\\
The contamination analysis adopted the top-hat aperture model with circular photometric apertures of 47" and 75"; these are effective contamination apertures rather than instrumental PSF sizes or full telescope fields of view. The 47" aperture corresponds to the nominal Mauve photometric aperture, while the 75" radius is adopted as a conservative configuration to account for the source-identification procedure, which may require searching for the target out to 75" \citep{2026RASTI...5ag018M}.\\
The contamination estimates presented in this section include only sources resolved in the input catalog. Companions closer to the target than the angular resolution of Gaia DR3, such as close binaries, are not listed as separate sources and are blended into the target photometry; PHOTO-CAT can therefore neither identify nor correct for them, and the reported values should be regarded as lower limits with respect to this population (Sect.~\ref{sec:lim_catalog}).
\subsection{Global contamination statistics}\label{sec:global_contamination}
For the full all-sky sample of 3.088.117 targets, PHOTO-CAT identifies the following for the 47" FoV:
\begin{itemize}
    \item 2.039.232 stars with contaminants
    \item 1.048.885 stars without contaminants
    \item 8.439.154 contaminants
\end{itemize}
This means that approximately 66\% of stars in the sample have at least one neighboring source within 47" that is bright enough ($\Delta$G $\leq$ 5) to contribute measurable flux contamination.
When the aperture is increased to 75", the number of contaminated stars and contaminants rises:
\begin{itemize}
    \item 2.531.920 stars with contaminants
    \item 556.197 stars without contaminants
    \item 21.137.161 contaminants
\end{itemize}
In this case, roughly 82\% of targets exhibit at least one contaminating neighbor. The number of contaminants increases by a factor of 2.50 almost exactly with the searching area. This behavior is consistent with the expectation for a nearly uniform projected density of contaminating sources.
\subsection{Distribution of contaminating sources}\label{sec:Distribution_of_contaminating_sources}
\begin{figure}
    \includegraphics[width=\hsize]{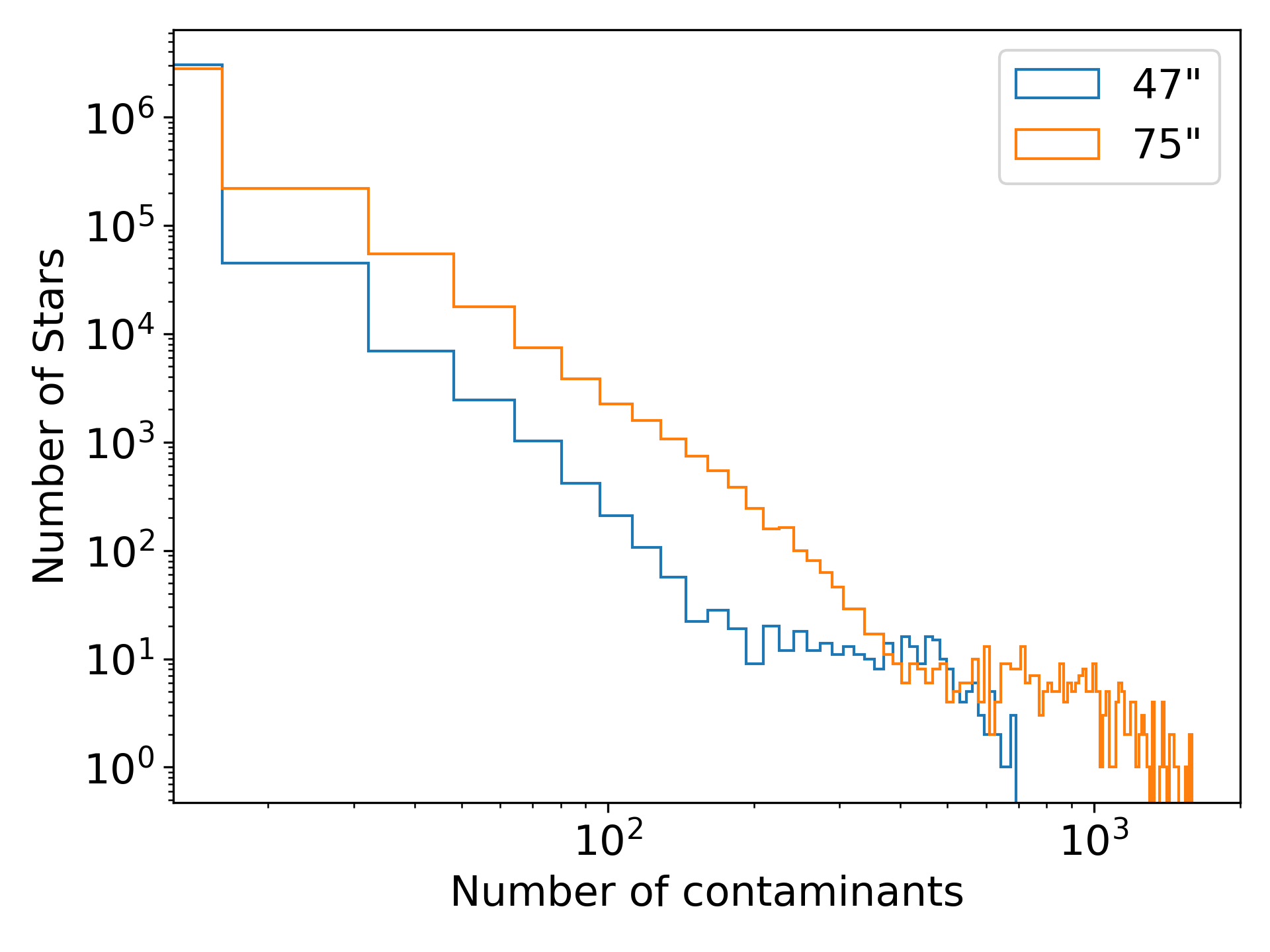}
    \caption{Distribution of the number of contaminating sources per target. The blue histogram shows targets within 47" FoV, while the orange histogram corresponds to a 75" FoV.}
    \label{fig:hist_contaminants}
\end{figure}

\begin{figure}
    \includegraphics[width=\hsize]{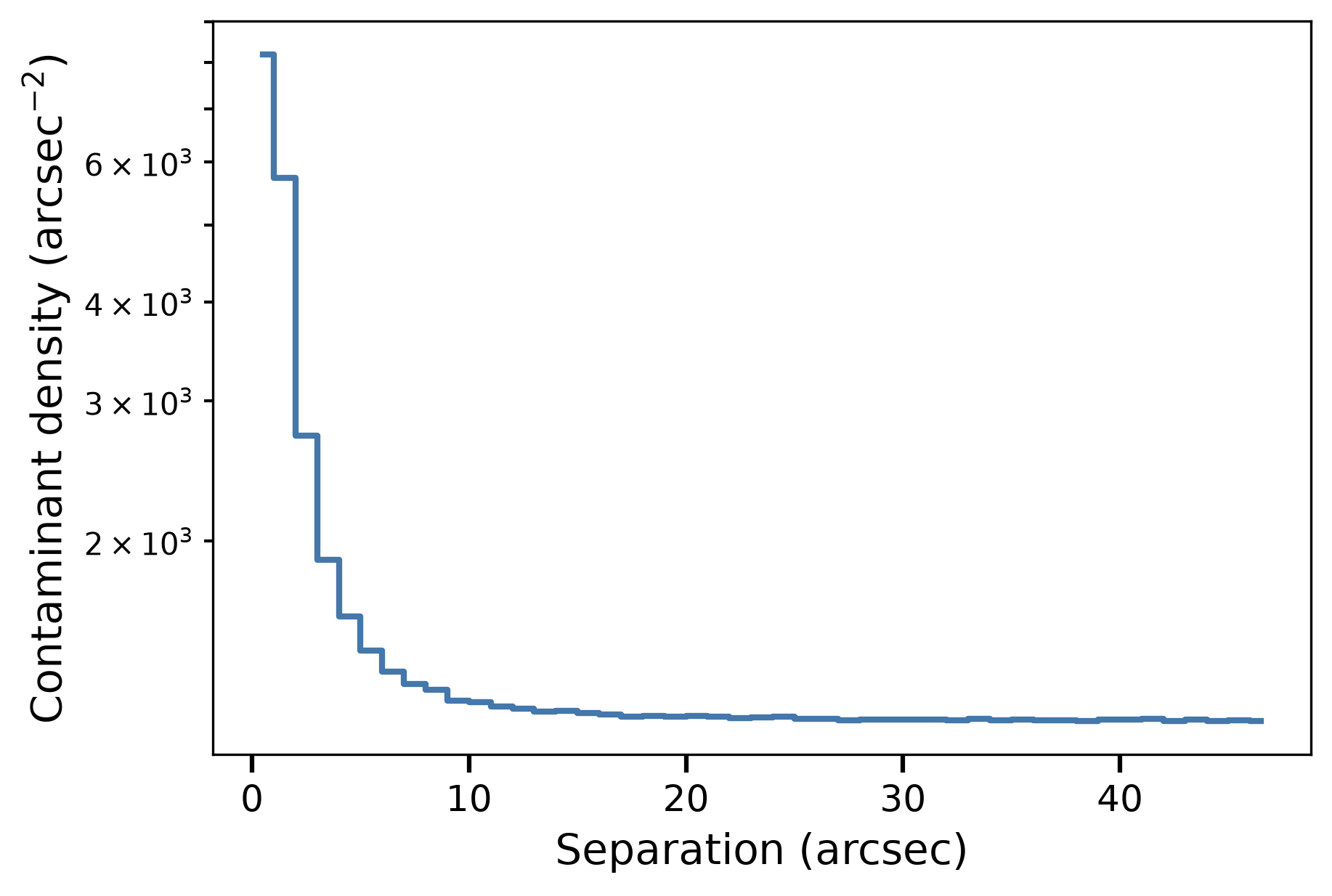}
    \caption{Surface density of contaminating sources vs angular separation from the target star for the 47" aperture and a magnitude contrast threshold of $\Delta G = 5$. The contaminant counts are normalized by the area of each 1-arcsec annulus, removing the geometric increase in searchable area with radius.}
    \label{fig:sep_distribution}
\end{figure}

The distribution of the number of contaminants per target is shown in Fig.~\ref{fig:hist_contaminants}. While most stars have few nearby companions, a non-negligible fraction show significantly larger contaminant counts, with a pronounced tail toward highly crowded environments, most notably along the Galactic plane. The 47" FoV distribution (blue) extends to $\sim$700 contaminants per target, while the 75" FoV (orange) reaches $\sim$1600, consistent with the larger search area sampling a proportionally larger number of stars at a given surface density.\\
The surface density of contaminating sources as a function of angular separation (Fig.~\ref{fig:sep_distribution}), normalized by annular area $\pi(r_{\rm out}^2-r_{\rm in}^2)$ for the 47" aperture and $\Delta G=5$, decreases toward small separations before flattening beyond $\sim$10". The excess at small angular scales indicates that contaminating sources are not distributed as a uniform random field around targets, consistent with close stellar companions, unresolved multiples, and stellar associations. At larger separations the density approaches the average local stellar density, where the contamination environment is dominated by the general field population rather than physical association.\\
A fraction of this apparent excess at small separations may be attributable to known processing artifacts in the Gaia data pipeline \citep{torra2021gaia}. Although the pipeline includes procedures to identify and merge duplicate detections, residual spurious sources can still persist \citep{torra2021gaia, fabricius2021gaia, babusiaux2023gaia}.

\subsection{Sky-map of contamination}\label{sec:sky-map}
\begin{figure}
    \includegraphics[width=\hsize]{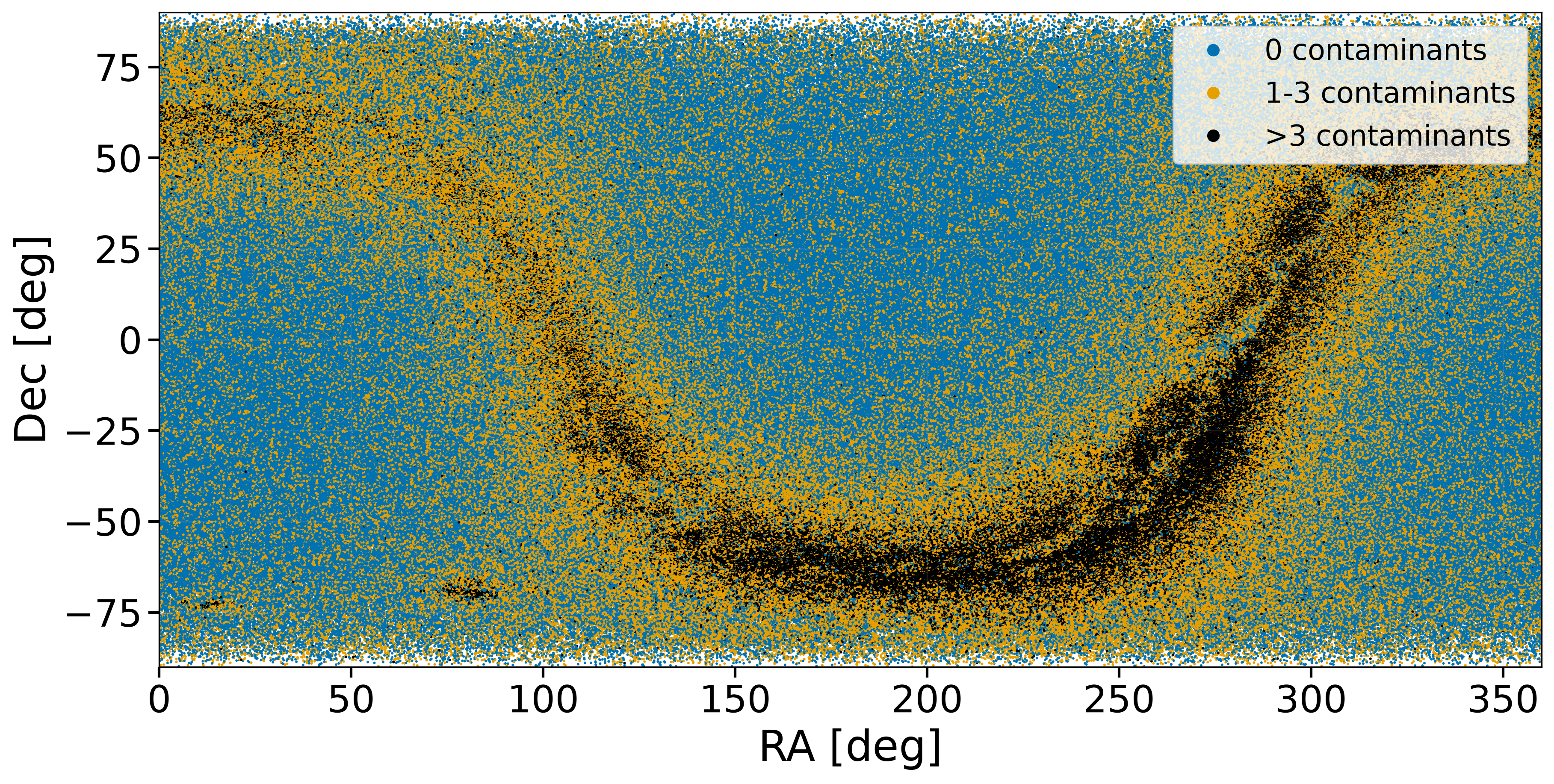}
    \caption{Sky map of stellar contamination for targets with $G \leq 12$ and a FoV of 47". Each point represents a target star. Blue points indicate stars without contaminating sources, orange points correspond to targets with 1--3 nearby contaminating sources, and black points mark targets with more than three contaminants.}
    \label{fig:sky_map}
\end{figure}

Fig.~\ref{fig:sky_map} shows the all-sky contamination map for the 47" FoV, $\Delta G=5$ configuration, classifying targets as: no contaminants (blue), 1--3 contaminants (orange), and more than 3 contaminants (black).\\
The dominant structure follows the Galactic plane, where high stellar densities drive the highest contamination levels; variations along the plane likely reflect spatially varying extinction and changes in the underlying stellar population. A clear localized overdensity also appears near the Large Magellanic Cloud (RA $\sim$80°, Dec $\sim$-70°), consistent with its high stellar density.\\
Contamination is thus not uniform across the sky but closely follows the large-scale structure of the Milky Way and nearby dense stellar systems. In practice, maps of this kind allow a mission to identify, before scheduling, the sky regions and individual targets that require closer inspection: targets with no contaminants can be observed without further checks, whereas targets in the highest contamination class can be inspected through their per-target PHOTO-CAT outputs, which list the position, separation, and flux of each contaminating source, before being included in the observing plan.

\subsection{Application to the Ariel Mission Candidate Sample}\label{sec:ariel_application}
\begin{figure*}
    \centering
    \subfloat[]{
        \includegraphics[width=0.43\hsize]{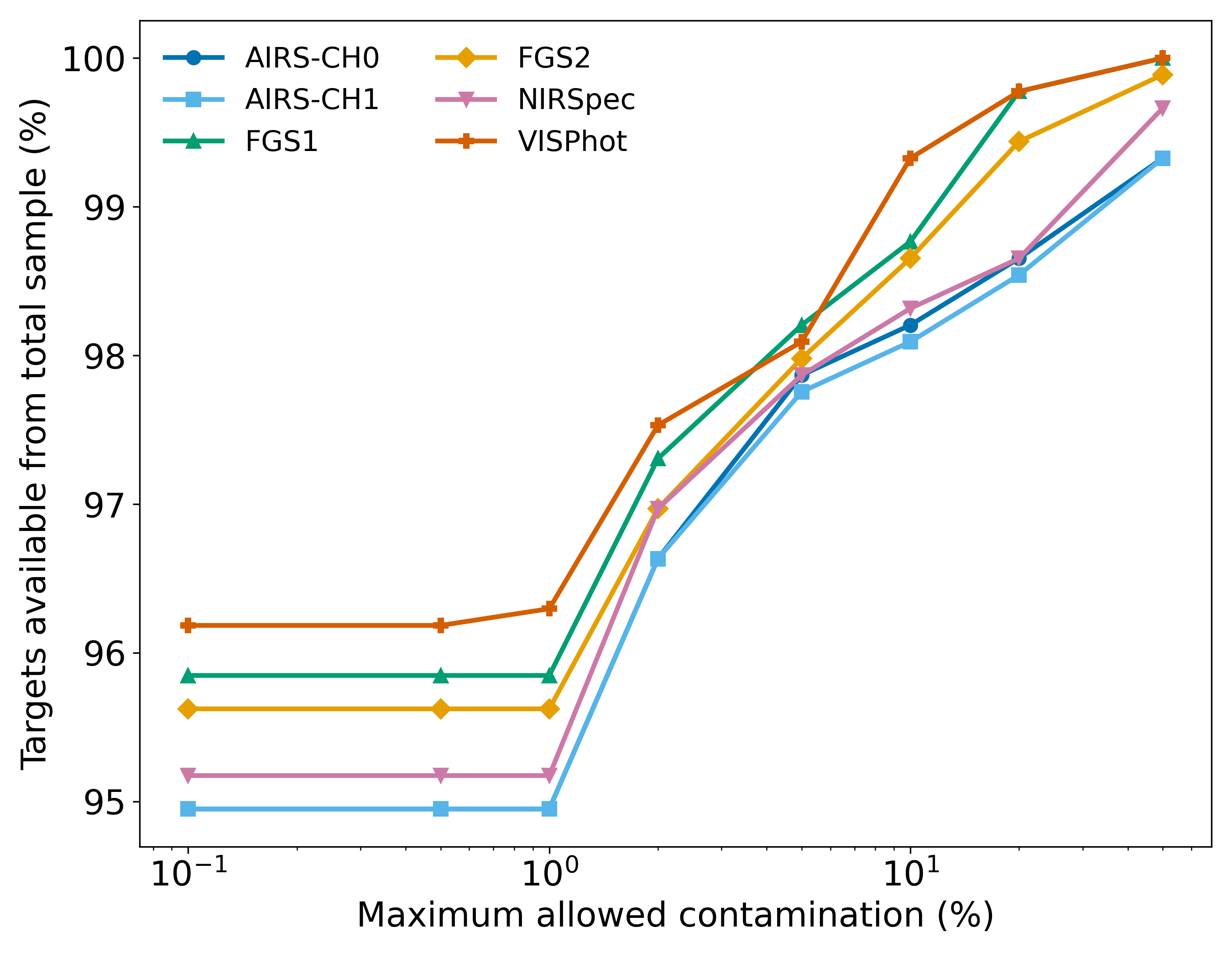}
        \label{fig:Ariel_filter_comparison_FoV3}
    }
    \subfloat[]{
        \includegraphics[width=0.43\hsize]{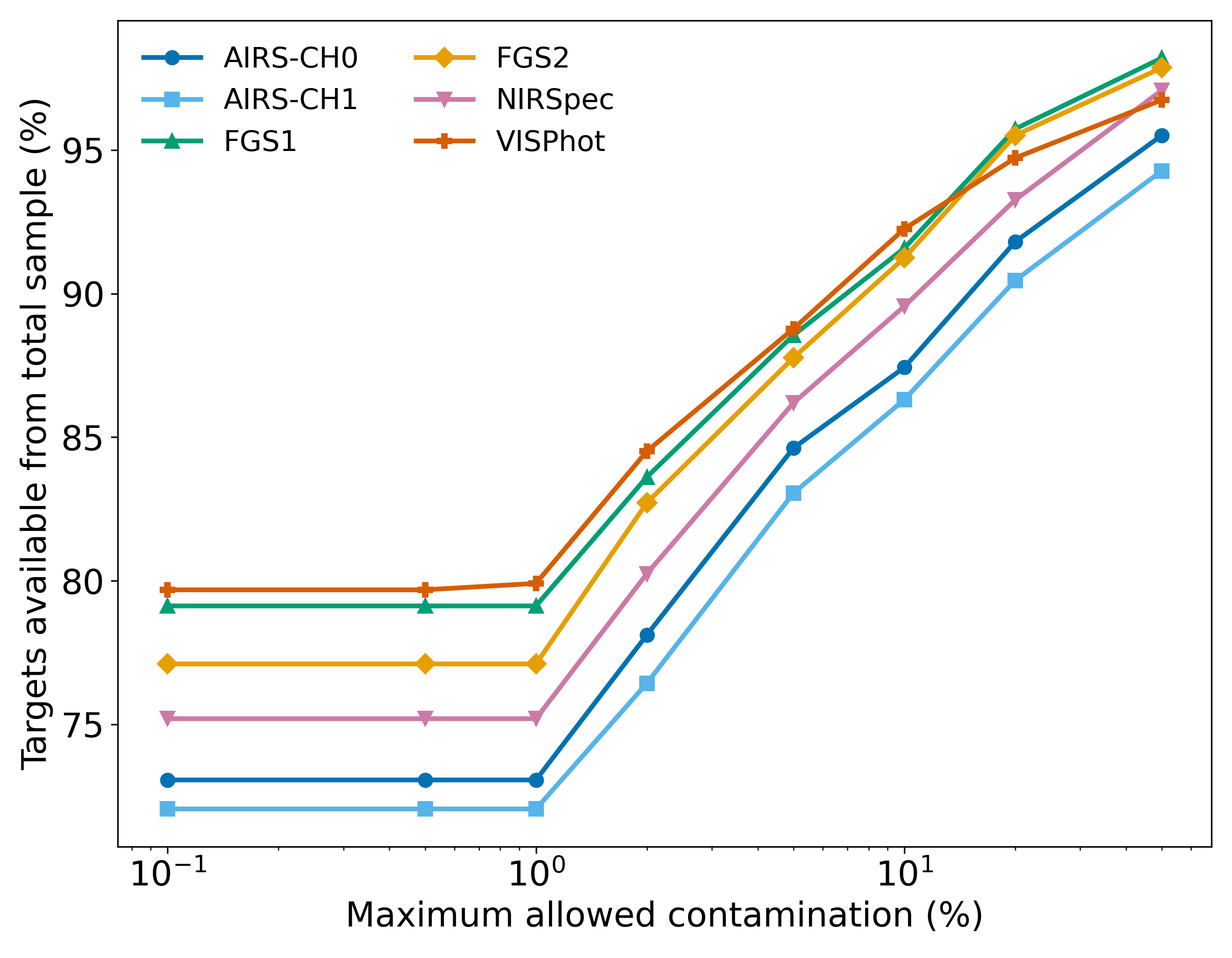}
        \label{fig:Ariel_filter_comparison_FoV13}
    }

    \caption{Comparison of the contamination impact among the six Ariel instrumental filters. 
    (a) Fraction of retained Ariel targets as a function of the maximum allowed contaminating flux contribution for the $3.2$" contamination radius, corresponding to half of the nominal slit width. 
    (b) Same as panel (a), but for the conservative $13.2$" contamination radius, corresponding to half of the nominal slit length. 
    The different curves show the chromatic dependence of contamination for the AIRS-CH0, AIRS-CH1, FGS1, FGS2, NIRSpec, and VISPhot channels.}

    \label{fig:Ariel_filter_comparison}
\end{figure*}

\begin{figure*}
    \centering
    \subfloat[]{
        \includegraphics[width=0.4\hsize]{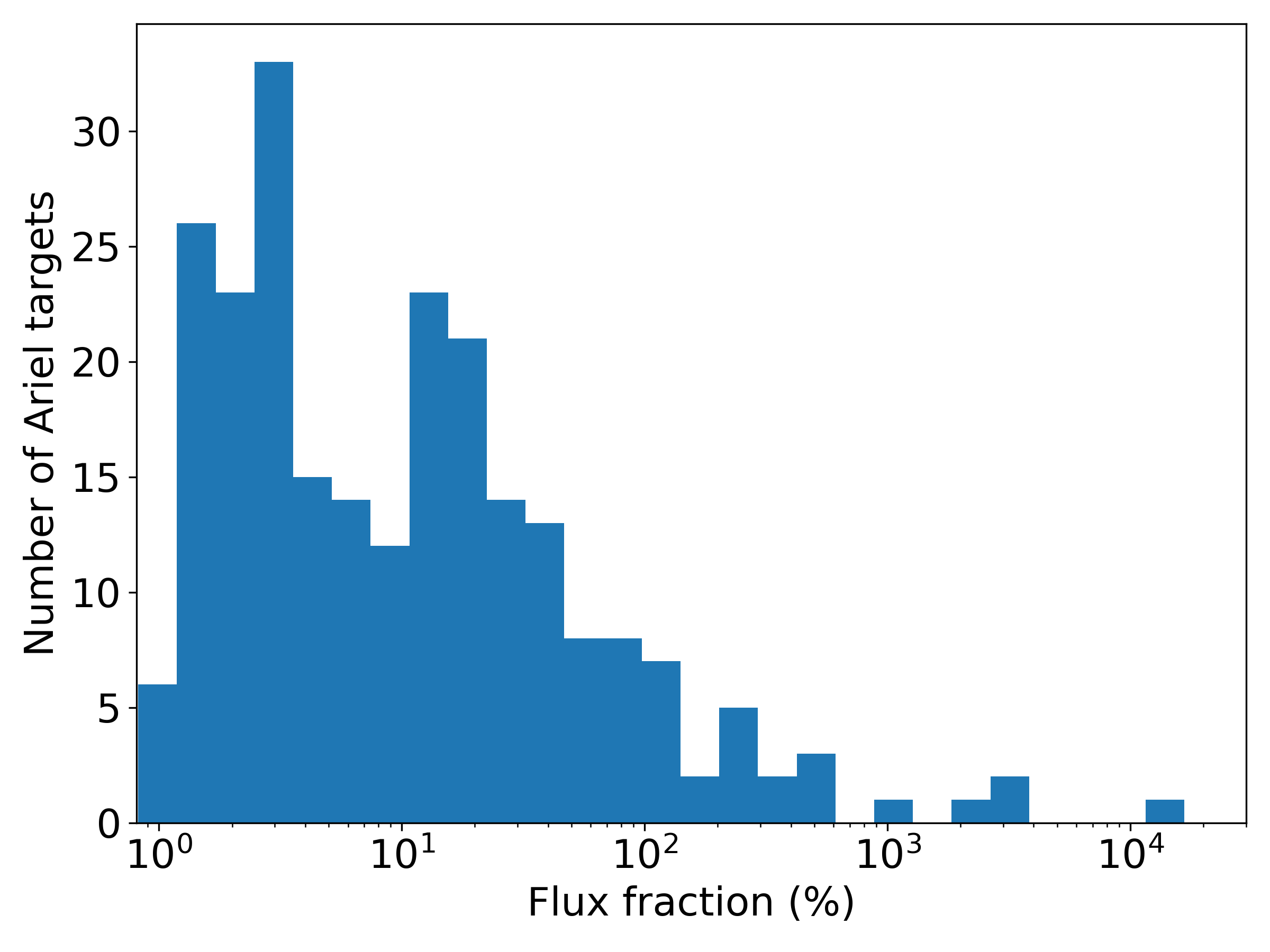}
        \label{fig:Ariel_contamination_distribution_CH0}
    }
    \subfloat[]{
        \includegraphics[width=0.42\hsize]{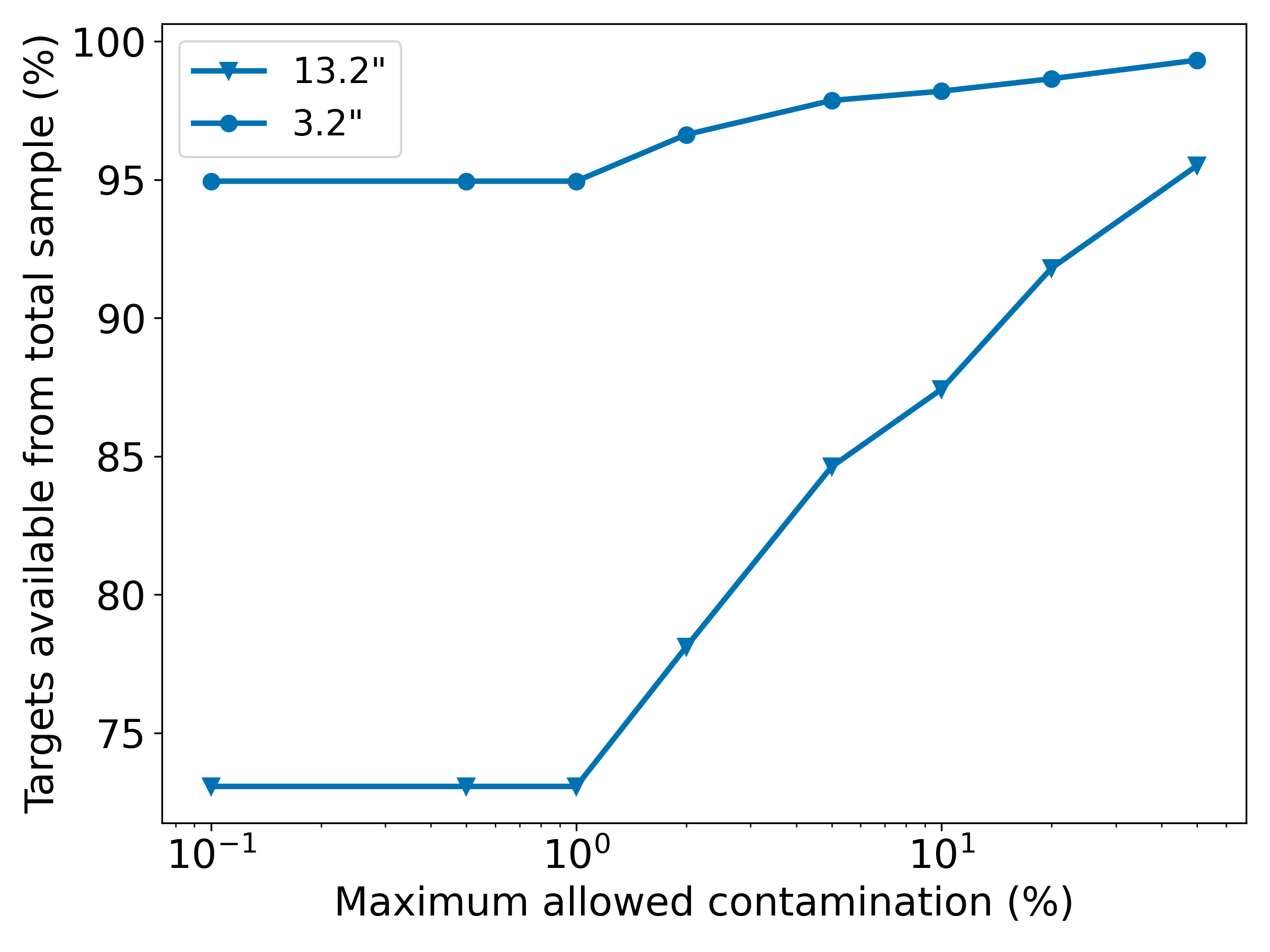}
        \label{fig:Ariel_target_selection_comparison_CH0}
    }

    \caption{Application of PHOTO-CAT to the Ariel AIRS-CH0 channel. 
    (a) Distribution of the contaminating flux fraction for the AIRS-CH0 band considering all 891 Ariel targets within a contamination radius of $13.2$".  
    (b) Fraction of retained Ariel targets as a function of the maximum allowed contamination threshold for two contamination radii: $13.2$" (downward triangle), corresponding to the conservative slit-length approximation, and $3.2$" (circle), corresponding to the slit half-width.}

    \label{fig:Ariel_CH0}
\end{figure*}

\begin{figure*}
    \centering
    \subfloat[]{
        \includegraphics[width=0.4\hsize]{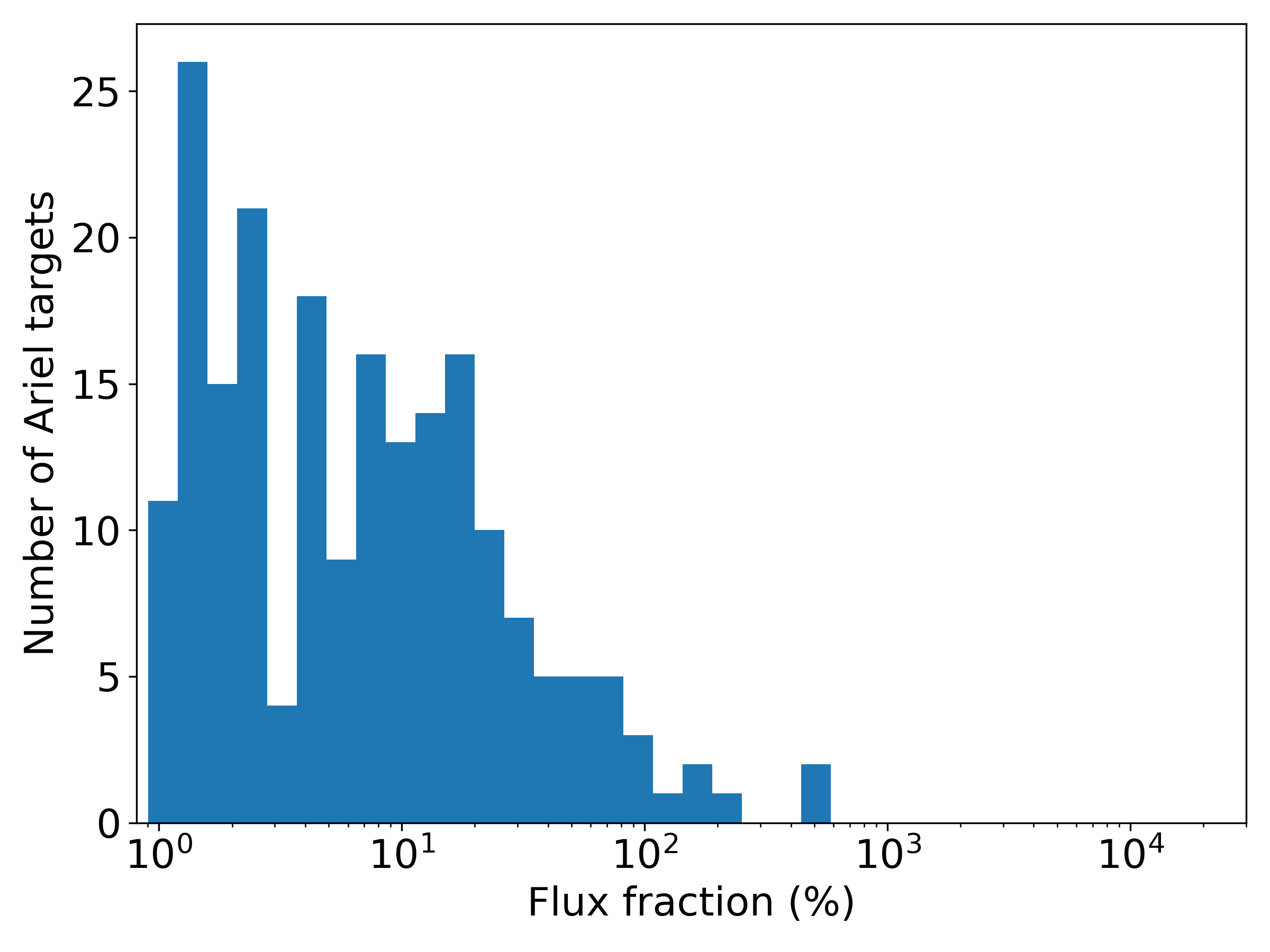}
        \label{fig:Ariel_contamination_distribution_FGS2}
    }
    \subfloat[]{
        \includegraphics[width=0.42\hsize]{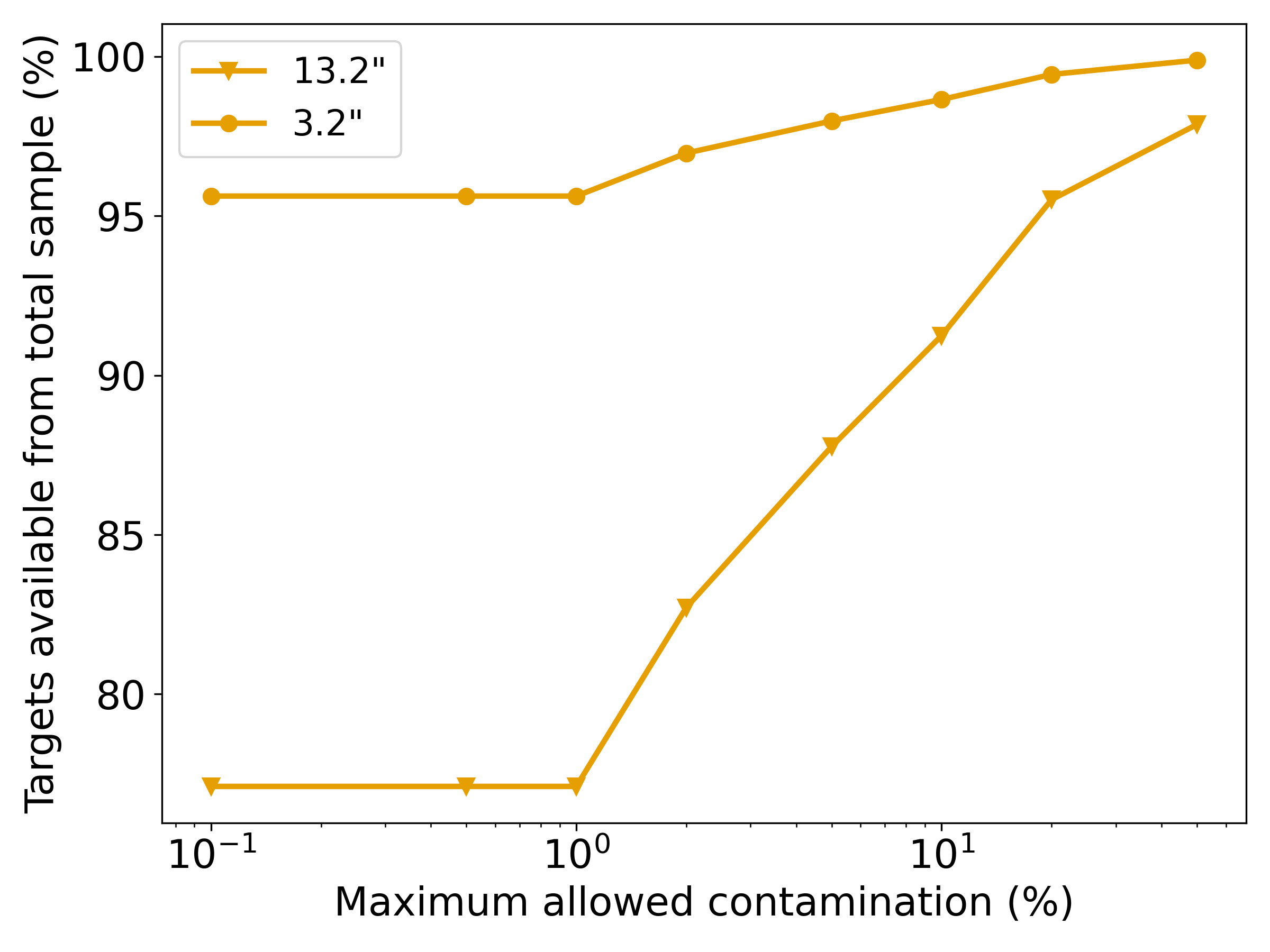}
        \label{fig:Ariel_target_selection_comparison_FGS2}
    }

    \caption{Application of PHOTO-CAT to the Ariel FGS2 channel. 
    (a) Distribution of the contaminating flux fraction for the FGS2 band within the $13.2$" contamination radius. 
    (b) Fraction of retained Ariel targets as a function of the maximum allowed contamination threshold for two contamination radii: $13.2$" (downward triangle), corresponding to the conservative slit-length approximation, and $3.2$" (circle), corresponding to the slit half-width.}

    \label{fig:Ariel_FGS2}
\end{figure*}

To demonstrate PHOTO-CAT's applicability to a real target-selection problem, we applied it to 891 confirmed exoplanet host stars of the Ariel Mission Candidate Sample (MCS) released in May 2026, selected for atmospheric characterization following the methodology of \cite{edwards2022ARIEL, edwards2019updated, mugnai2020arielrad}.\\
Ariel observations use a rectangular slit ($26.4'' \times 6.4''$; \citealt{nascimbeni2017ariel}) rather than a circular aperture. We adopted a first-order circular approximation with two radii: $13.2$" (half the slit length, a conservative bound including all sources within the maximum slit extent) and $3.2$" (half the slit width, isolating contaminants that are harder to avoid through slit-orientation optimization). Both cases use the top-hat model.\\
Unlike the all-sky analysis, no limiting magnitude for flux contamination was imposed on neighbors: all Gaia DR3 sources within each search radius were considered, with contaminants identified at $\Delta G \leq 5$. 
Before performing the SED-based conversion, we derived the stellar parameters required by the atmosphere models for the MCS sources whenever possible. Existing Gaia PHOENIX parameters were retained. For sources without a $T_{\rm eff}$ estimate, we constructed an empirical $BP-RP$--$T_{\rm eff}$ relation from the subset of Gaia sources with valid PHOENIX parameters, selecting stars with $A_0<0.05$, $|[{\rm M/H}]|<0.20$, and $\log g>4.0$. This selection was intended to minimize extinction and metallicity effects and to preferentially sample main-sequence stars. The relation was constructed by binning the calibration sources in $BP-RP$, computing the median $T_{\rm eff}$ in each bin, and applying a Piecewise Cubic Hermite Interpolating Polynomial \citep[PCHIP]{fritsch1980monotone} interpolation within the calibrated color range. A linear extrapolation was allowed within $0.5$ mag beyond the calibrated color boundaries.\\
For sources without $\log g$, we constructed an analogous $T_{\rm eff}$--$\log g$ relation from the same calibration population, again using the median $\log g$ in temperature bins followed by PCHIP interpolation. A linear extrapolation was permitted within a maximum temperature tolerance of $500$ K outside the calibrated range. When no metallicity measurement was available, solar metallicity was assumed, $[{\rm M/H}]=0$. These estimates were used solely to provide representative stellar parameters required for the construction of the model-atmosphere SEDs in this Ariel analysis. The parameter-completion procedure is therefore specific to the MCS preparation and is not part of the PHOTO-CAT band-conversion algorithm itself.\\
For sources lacking one or both Gaia BP/RP measurements, the $BP-RP$ color was estimated from stars with similar photometry. As a reference sample, we used all sources of the same input catalog with valid $G$, $BP$, and $RP$ photometry. For each source with incomplete photometry, we selected the 100 reference stars whose magnitudes are closest to those available for the source: $G$ alone when both BP and RP are missing, $G$ and $RP$ when only BP is missing, and $G$ and $BP$ when only RP is missing. The median $BP-RP$ color of these reference stars was adopted as the color of the source, and their spread as its uncertainty. The number of reference stars is a compromise between the statistical noise of small samples and the loss of locality for large ones; the results change by less than 0.01 mag for values between 30 and 200. Withholding 20\% of the reference stars and inferring their colors with the same procedure, the inferred colors reproduce the measured ones with an rms of 0.46, 0.29, and 0.24 mag when only $G$, $G$ and $RP$, or $G$ and $BP$ are available, respectively, with negligible bias ($\lesssim0.04$ mag). The inferred colors were then used to derive $T_{\rm eff}$ and $\log g$ through the relations described above. Because the inferred color represents the typical color of stars with similar photometry rather than a measurement, this procedure provides a statistical estimate of the contamination, appropriate for the population-level analysis presented here, but it is not reliable for individual sources.\\
Overall, among the 113.706 MCS sources, 30.29\% already had complete stellar parameters, 55.34\% had available $BP-RP$ colors but required parameter estimation, and 14.36\% required a statistical recovery of the $BP-RP$ color before the stellar parameters could be derived.\\
Contaminant contributions were converted from Gaia photometry into six Ariel instrumental bands (AIRS-CH0, AIRS-CH1, FGS1, FGS2, NIRSpec, VISPhot) using the SED-based band-conversion method of Sect. \ref{sec:atmosphere_conversion}, with transmission curves from the EXOSim2 instrument model \citep{mugnai2025exosim}.\\
For the conservative $13.2$" configuration, the number of targets with at least one selected contaminant varies from a minimum of 181/891 (20.31\%) in VISPhot to a maximum of 249/891 (27.95\%) in AIRS-CH1. A complete summary of the contaminated and uncontaminated targets across all six Ariel channels and both adopted radii is provided in Tab.~\ref{tab:contamination_ariel_channels}.\\
Figures~\ref{fig:Ariel_filter_comparison_FoV3} and \ref{fig:Ariel_filter_comparison_FoV13} compare the retained target fraction across the six filters for both radii.
The infrared channels are particularly sensitive to contamination in terms of contaminating flux. These large values arise because sources that are relatively faint in the optical can contribute substantially more flux at infrared wavelengths, either because they are highly reddened or because they are intrinsically red, such as red giants or M dwarfs. This illustrates why contamination must be evaluated directly in the instrumental band rather than from optical magnitudes alone.\\
Fig.~\ref{fig:Ariel_contamination_distribution_CH0} and Fig.~\ref{fig:Ariel_contamination_distribution_FGS2} show the flux contamination distributions for AIRS-CH0 and FGS2. The mean contaminating contribution decreases from 39.2\% to 8.1\%. For FGS2 the mean contamination decreases from 4.71\% to 0.42\%.\\
Fig.~\ref{fig:Ariel_target_selection_comparison_CH0} and Fig.~\ref{fig:Ariel_target_selection_comparison_FGS2} show the retained-target fraction as a function of the allowed contamination threshold for the two channels and aperture geometries. At a $1\%$ flux contamination threshold, 73.06\% of targets are retained in AIRS-CH0 for the $13.2$" configuration, compared with 94.95\% for the $3.2$" configuration. For FGS2, the corresponding fractions are 77.10\% and 95.62\%. The comparison between the two radii therefore distinguishes contamination arising from the broader target environment from contamination that may be mitigated through observing-geometry optimization.\\
These outputs can be translated directly into observing-strategy decisions. Since the $3.2$" region is enclosed by the slit for any orientation, whereas sources between $3.2$" and $13.2$" fall inside the slit only for some position angles and pointing offsets, the targets can be grouped into three classes. (i) Targets with no selected contaminant within $13.2$" can be observed with any slit orientation. (ii) Targets with selected contaminants only between $3.2$" and $13.2$" can be observed with a slit position angle, or a small pointing offset of the target within the slit, that excludes them; in AIRS-CH0, this class includes 195 of the 891 targets (21.9\%; Table~\ref{tab:contamination_ariel_channels}), whose contamination could in principle be avoided through slit-orientation or pointing-offset optimization, within the range of position angles and offsets allowed by the observing constraints. (iii) Targets with selected contaminants within $3.2$" (45 targets, 5.05\%, in AIRS-CH0) cannot be cleaned by any orientation, and an offset large enough to exclude these contaminants would displace the target from the slit center, at the cost of part of its own flux: for these targets, the band-specific contamination estimated by PHOTO-CAT should be compared with the precision and accuracy requirements of the planned observations, and the target either deprioritized or flagged for a contamination correction in the data analysis. The retained-target curves of Figs.~\ref{fig:Ariel_target_selection_comparison_CH0} and \ref{fig:Ariel_target_selection_comparison_FGS2} provide the corresponding trade-off between the adopted contamination threshold and the size of the usable sample.\\
Overall, the majority of Ariel targets remain uncontaminated in all channels, with $72.05$--$79.69\%$ of targets showing no selected contaminant within $13.2$" and $94.95$--$96.18\%$ within $3.2$". Nevertheless, a non-negligible fraction of targets is affected by contamination, with the incidence and contaminating flux strongly dependent on wavelength. These target-by-target, instrumental-band estimates demonstrate the importance of evaluating contamination in the actual observing band and provide a realistic basis for target selection and scheduling for high-precision spectrophotometric observations.
\begin{table}
\centering
\caption{Summary of the number and fraction of Ariel MCS targets affected by at least one selected contaminant and of uncontaminated targets in each instrumental channel, for the two adopted circular-equivalent radius. The $13.2$" radius provides a conservative estimate of the contamination environment, whereas the $3.2$" radius represents the narrower slit width.}
\label{tab:contamination_ariel_channels}
\resizebox{\columnwidth}{!}{%
\begin{tabular}{cccc}
\hline
Ariel channel             & FoV (arcsec) & Contaminated  & Non contaminated \\ \hline
\multirow{2}{*}{AIRS-CH0} & 13.2         & 240 (26.94\%) & 651 (73.06\%)    \\
                          & 3.2          & 45 (5.05\%)   & 846 (94.95\%)    \\
\multirow{2}{*}{AIRS-CH1} & 13.2         & 249 (27.95\%) & 642 (72.05\%)    \\
                          & 3.2          & 45 (5.05\%)   & 846 (94.95\%)    \\
\multirow{2}{*}{FGS1}     & 13.2         & 186 (20.88\%) & 705 (79.12\%)    \\
                          & 3.2          & 37 (4.15\%)   & 854 (95.85\%)    \\
\multirow{2}{*}{FGS2}     & 13.2         & 204 (22.90\%) & 687 (77.10\%)    \\
                          & 3.2          & 39 (4.38\%)   & 852 (95.62\%)    \\
\multirow{2}{*}{NIRSpec}  & 13.2         & 221 (24.80\%) & 670 (75.20\%)    \\
                          & 3.2          & 43 (4.83\%)   & 848 (95.17\%)    \\
\multirow{2}{*}{VISphot}  & 13.2         & 181 (20.31\%) & 710 (79.69\%)    \\
                          & 3.2          & 34 (3.82\%)   & 857 (96.18\%)   
\end{tabular}%
}
\end{table}
\section{Performance and Operational Considerations}\label{sec:Performance_and_Operational_Considerations}
This section evaluates PHOTO-CAT's computational performance and scalability (Sect.~\ref{sec:Performance_Evaluation_and_Scalability}), the design choices enabling efficient execution (Sect.~\ref{sec:Efficiency_Considerations}).
\subsection{Performance Evaluation and Scalability}\label{sec:Performance_Evaluation_and_Scalability}
PHOTO-CAT's computational profile is dominated by a single stage: construction of the stellar neighborhood index (converting coordinates to 3D unit vectors, building the KDTree, streaming neighbor lists in CSR format), the only phase requiring an exhaustive pass over the full catalog.\\
Using a Gaia DR3 dataset within $-47^\circ \leq Dec \leq +33^\circ$ with $G \leq 17$, corresponding to the preliminary Mauve satellite field of regard (FoR) \citep[][]{saba2025mauvesim} ($\sim$95 million sources after quality filtering), index construction requires $\sim$30 minutes (Intel i9-class CPU, $\sim$28 GB RAM), with $\sim$62 GB of output on disk. An additional full-sky benchmark on $\sim$157 million sources (47" radius, $\Delta G=5$) required $\sim$50 minutes on the same hardware. Table~\ref{tab:performance} summarizes both benchmarks. This is a one-time cost: once built, the index requires no repeated geometric computation for subsequent analyses.
\begin{table}
\caption{Performance benchmarks of PHOTO-CAT for different catalog configurations.}
\label{tab:performance}
\centering
\begin{tabular}{lcc}
\hline
Parameter & Mauve FoR benchmark & Full-sky benchmark \\
\hline
Catalog & Gaia DR3 & Gaia DR3 \\
Sky coverage & $-47^\circ \leq Dec \leq +33^\circ$ & Full sky \\
Number of sources & $\sim95$ million & $\sim157$ million \\
Magnitude limit & $G \leq 17$ & $G \leq 17$ \\
Search radius & 47" & 47" \\
Magnitude threshold & $\Delta G = 5$ & $\Delta G = 5$ \\
Index construction time & $\sim30$ min & $\sim50$ min \\
Hardware & \multicolumn{2}{c}{Intel i9-class CPU, 28 GB RAM} \\
Index storage & $\sim62$ GB & $\sim80$ GB \\
\hline
\end{tabular}
\end{table}
Tasks such as loading a target subset, retrieving its neighborhood via CSR pointer arithmetic, and evaluating contamination metrics require only a few minutes even for $10^4-10^5$ stars, since memory-mapped NumPy arrays touch only the relevant slices without a global scan of the neighbor list; subsets up to $\sim10^6$ targets remain feasible without recomputing the underlying geometry.\\
This division between a one-time preprocessing stage and lightweight queries is central to the tool's design, enabling rapid exploration of parameter spaces (FoV radius, magnitude thresholds, target lists) while keeping the computational burden concentrated in a single, resumable pass.
\subsection{Efficiency Considerations}\label{sec:Efficiency_Considerations}
Although Python incurs interpreter-level overhead, most computationally intensive operations in PHOTO-CAT are delegated to compiled libraries (NumPy, SciPy, spatial indexing routines), so critical sections execute at near-native speed while Python serves as a high-level orchestration layer.\\
KD-tree indexing reduces neighbor-search complexity from $O(N^2)$ to $\sim O(N\log N)$, and compact 64-bit integer identifiers (Sect.~\ref{sec:Compact_ID_Mapping}) improve cache locality and reduce memory overhead independently of whether the code is compiled or interpreted.
\section{Limitations and Future Developments}\label{sec:Tool_Limitations}
This section summarizes the main limitations of the current implementation of PHOTO-CAT, grouped according to their origin: the input catalog (Sect.~\ref{sec:lim_catalog}), the photometric band conversion (Sect.~\ref{sec:lim_conversion}), the aperture and PSF models (Sect.~\ref{sec:lim_psf}), and the computational resources (Sect.~\ref{sec:lim_computing}). The planned developments addressing these limitations are outlined in Sect.~\ref{sec:future_developments}.
\subsection{Input catalog}\label{sec:lim_catalog}
PHOTO-CAT can only account for the sources listed in the input catalog, and its results therefore inherit the completeness, angular resolution, astrometric quality, and photometric quality of that catalog. Three consequences follow.\
First, companions closer to the target than the angular resolution of the underlying survey, which is sub-arcsecond for Gaia DR3 \citep{fabricius2021gaia}, are not listed as separate sources: their flux is blended into the catalog photometry of the target itself. Such unresolved companions, and in particular close binaries, are arguably the most pernicious contaminants, since no choice of aperture or observing geometry can separate them from the target. PHOTO-CAT can neither identify nor correct for them, and the contamination estimates it provides should therefore be regarded as lower limits with respect to this population. Their identification requires dedicated high-angular-resolution observations, such as speckle or adaptive-optics imaging \citep[e.g.][]{2015ApJ...805...16C, 2021AJ....162...75L}.\
Second, all catalog sources are treated as point sources with a defined position and integrated flux. The accuracy of the results therefore depends on the catalog completeness, which for Gaia DR3 is high for faint and intermediate-magnitude stars but limited for the brightest ones \citep[][]{brown2020gaia}. The point-source approximation also fails for extended objects (galaxies, nebulae, crowded associations), whose morphology cannot be described by a single position and scalar flux; although these constitute a small fraction of the catalog, they can affect the local background in ways not captured by the current framework.\
Third, the input catalog is treated as a static snapshot of the sky. Stellar variability, orbital motion in binary systems, or large proper motions over multi-year campaigns can alter the true source configuration relative to the catalog epoch. This effect is small for most stars but may be non-negligible for rapidly moving nearby dwarfs or long-period binaries.
\subsection{Photometric band conversion}\label{sec:lim_conversion}
Each of the band-conversion methods of Sect.~\ref{sec:bandpass_conversion} introduces its own sources of uncertainty. The empirical Gaia transformations are restricted to the color ranges, stellar populations, and photometric systems for which they were calibrated, and therefore do not provide a general solution for arbitrary passbands. The atmosphere-based conversion provides a physically detailed description of the stellar spectrum, but depends on the coverage and assumptions of the adopted model grids and on the accuracy of the stellar parameters used to select and interpolate the models; uncertainties in stellar colors, atmospheric parameters, or model assumptions propagate directly into the estimated contamination. These limitations are most relevant when extrapolating from broad-band optical photometry to infrared bands, where cool companions or background sources may contribute more flux than their optical brightness suggests.
\subsection{Aperture and PSF models}\label{sec:lim_psf}
The two aperture models are simplified descriptions of the true instrumental response. The top-hat model assumes a uniform flux contribution within a fixed circular radius and neglects the extended wings of any real PSF. The Gaussian PSF model weights the contribution of each contaminant continuously with distance, but remains an idealized, circularly symmetric profile defined by a single FWHM: it does not capture PSF asymmetries, diffraction structures, detector effects, or wavelength-dependent variations. The present analytical approach may therefore underestimate or overestimate contamination when the actual instrumental footprint is strongly non-circular, as for a slit, or when extended or non-Gaussian PSF wings are important. Mission-specific contamination estimates require calibrated instrumental PSF and aperture models, which are not yet implemented in PHOTO-CAT. PHOTO-CAT is accordingly intended, in its current form, as a contamination risk-assessment and target-screening tool rather than a substitute for mission-specific photometric extraction pipelines.
\subsection{Computational resources}\label{sec:lim_computing}
The CSR-based neighbor index is highly efficient compared to recomputing neighborhoods on demand, but carries a substantial storage cost ($\sim$62 GB for the Mauve field of regard and $\sim$80 GB for the full-sky configuration; Table~\ref{tab:performance}), which imposes a practical constraint on smaller machines.
\subsection{Future developments}\label{sec:future_developments}
Future developments of PHOTO-CAT will address the limitations above along two main directions.\
The first concerns the physical fidelity of the contamination estimates. The modular design of the tool allows calibrated, mission-specific PSF models to be integrated without modifying the core contamination framework. Such models would capture non-Gaussian wings, diffraction structures, and spatial or wavelength-dependent variations that the current top-hat and Gaussian approximations cannot reproduce, and would extend the existing wavelength-dependent framework (Sect.~\ref{sec:bandpass_conversion}) to a PSF characterization that varies with wavelength and detector position.\
The second concerns computational efficiency. A compression layer for the CSR neighbor index could reduce its storage footprint by a factor of three to six, while parallelization and algorithmic improvements could accelerate index construction and querying.\
Taken together, these developments aim to make the tool faster, more memory-efficient, and more broadly applicable, keeping pace with the growing size of astronomical catalogs and the increasingly demanding contamination requirements of next-generation surveys.
\section{Conclusions}\label{sec:Conclusions}
We presented PHOTO-CAT, a scalable, mission-agnostic tool for quantifying stellar flux contamination in wide-field photometric observations, combining a precomputed stellar neighborhood index, configurable top-hat or Gaussian PSF aperture models, and flexible wavelength-dependent flux conversion. The latter can be performed using empirical Gaia DR3 photometric relations or synthetic stellar-atmosphere spectra combined with arbitrary instrumental transmission curves (Sect.~\ref{sec:workflowstructure}).\\
Performance tests (Sect.~\ref{sec:Performance_Evaluation_and_Scalability}) show that PHOTO-CAT handles catalogs of tens to hundreds of millions of stars with predictable memory usage and tractable runtimes, thanks to the streaming CSR index and compact identifier mapping; the precomputed index enables rapid repeated queries for arbitrary target lists.\\
Applying PHOTO-CAT to an all-sky Gaia sample (G$\leq$12, 3.09 million targets, neighbors to G=17) shows that 66\% of stars have at least one contaminating neighbor ($\Delta G\leq5$) within a 47" aperture, rising to 82\% within 75". Contamination depends strongly on aperture geometry, target brightness, and local stellar density, peaking along the Galactic plane (Sect.~\ref{sec:global_contamination}--\ref{sec:sky-map}).\\
Applied to the Ariel Mission Candidate Sample (Sect.~\ref{sec:ariel_application}), PHOTO-CAT's wavelength-dependent estimates show that contamination varies substantially across instrumental channels, most severely in the infrared (AIRS-CH0, AIRS-CH1), confirming that band-specific contamination assessment should be considered for observing-geometry optimization (rotation or offset instrumentation).\\
These results demonstrate PHOTO-CAT's value for pre-observation risk assessment: supporting target selection and field planning at both survey-wide and individual-target levels, while remaining a screening tool that complements, rather than replaces, instrument-specific photometric extraction and calibration pipelines, and that accounts only for sources resolved in the input catalog, so that companions below the catalog angular resolution require dedicated high-resolution observations. Planned developments will reduce the index's memory footprint and incorporate calibrated, mission-specific PSF models to further improve the physical fidelity of contamination predictions as catalogs and survey requirements grow in scale and complexity (Sect.~\ref{sec:Tool_Limitations}).

\section*{Acknowledgements}
We thank the two anonymous referees for their helpful suggestions that improved the clarity and content of the paper.\\
G. M. and G. G. acknowledge support from the European Union - Next Generation EU through the grant n. 2022J7ZFRA - Exo-planetary Cloudy Atmospheres and Stellar High energy (Exo-CASH) funded by MUR - PRIN 2022 and the ASI-INAF agreement 2021-5-HH.2-2024.\\
G. G. acknowledges the support of Blue Skies Space Ltd. (\url{www.bssl.space}).\\
This research has made use of the SVO Filter Profile Service "Carlos Rodrigo", funded by MCIN/AEI/10.13039/501100011033/ through grant PID2023-146210NB-I00.\\
This work has made use of data from the European Space Agency (ESA) mission {\it Gaia} (\url{https://www.cosmos.esa.int/gaia}), processed by the {\it Gaia} Data Processing and Analysis Consortium (DPAC, \url{https://www.cosmos.esa.int/web/gaia/dpac/consortium}). Funding for the DPAC has been provided by national institutions, in particular the institutions participating in the {\it Gaia} Multilateral Agreement.
\section*{Data Availability}
The PHOTO-CAT tool presented in this work is publicly available at \url{https://github.com/Daddi14/photo-cat}.\\
The complete target-by-target outputs generated by PHOTO-CAT for the Ariel MCS analysis are publicly available through the associated Zenodo repository at \url{https://zenodo.org/records/22639037}.\\
This work has made use of data from the European Space Agency (ESA) mission
{\it Gaia} (\url{https://www.cosmos.esa.int/gaia}), processed by the {\it Gaia}
Data Processing and Analysis Consortium (DPAC,
\url{https://www.cosmos.esa.int/web/gaia/dpac/consortium}). Funding for the DPAC
has been provided by national institutions, in particular the institutions
participating in the {\it Gaia} Multilateral Agreement.
\section*{Conflict of Interest}
The authors declare no conflict of interest.



\bibliographystyle{rasti}
\bibliography{example} 

@ARTICLE{2026RASTI...5ag018M,
       author = {{Mauve Science Collaboration - Year 1:} and {Ag{\"u}eros}, Marcel A. and {Dixon}, Don and {Dong}, Chuanfei and {Duvvuri}, Girish M. and {Flanagan}, Patrick F. and {Johns-Krull}, Christopher M. and {Lu}, Hongpeng and {Maehara}, Hiroyuki and {Namekata}, Kosuke and {N{\'u}{\~n}ez}, Alejandro and {Pancino}, Elena and {Rani}, Sharmila and {Ravikumar}, Anusha and {Sigut}, T.~A.~A. and {Stassun}, Keivan G. and {Stewart}, Jamie J. and {Vida}, Kriszti{\'a}n and {Whelan}, Emma T. and {Wilcock}, Benjamin J. and {Razin}, Sharafina and {Saba}, Arianna and {Stotesbury}, Ian and {Tinetti}, Giovanna and {Tessenyi}, Marcell and {Tennyson}, Jonathan},
        title = "{The Preliminary Mauve Science Programme: science themes identified for the first year of operations}",
      journal = {RAS Techniques and Instruments},
         year = 2026,
        month = jan,
       volume = {5},
          eid = {rzag018},
        pages = {rzag018},
          doi = {10.1093/rasti/rzag018},
archivePrefix = {arXiv},
       eprint = {2512.16675},
 primaryClass = {astro-ph.SR},
       adsurl = {https://ui.adsabs.harvard.edu/abs/2026RASTI...5ag018M}
}

@inproceedings{tinetti2016science,
  title={The science of ARIEL (atmospheric remote-sensing infrared exoplanet large-survey)},
  author={Tinetti, Giovanna and Drossart, Pierre and Eccleston, Paul and Hartogh, Paul and Heske, A and Leconte, J and Micela, G and Ollivier, M and Pilbratt, G and Puig, L and others},
  booktitle={Space Telescopes and Instrumentation 2016: Optical, Infrared, and Millimeter Wave},
  volume={9904},
  pages={658--667},
  year={2016},
  organization={SPIE}
}

@article{tinetti2018chemical,
  title={A chemical survey of exoplanets with ARIEL},
  author={Tinetti, Giovanna and Drossart, Pierre and Eccleston, Paul and Hartogh, Paul and Heske, Astrid and Leconte, J{\'e}r{\'e}my and Micela, Giusi and Ollivier, Marc and Pilbratt, G{\"o}ran and Puig, Ludovic and others},
  journal={Experimental astronomy},
  volume={46},
  number={1},
  pages={135--209},
  year={2018},
  publisher={Springer}
}

@INPROCEEDINGS{2022EPSC...16.1114T,
       author = {{Tinetti}, Giovanna and {Eccleston}, Paul and {Lueftinger}, Theresa and {Salvignol}, Jean-Christophe and {Fahmy}, Salma and {Alves de Oliveira}, Caterina},
        title = "{Ariel: Enabling planetary science across light-years}",
    booktitle = {European Planetary Science Congress},
         year = 2022,
        month = sep,
          eid = {EPSC2022-1114},
        pages = {EPSC2022-1114},
          doi = {10.5194/epsc2022-1114},
archivePrefix = {arXiv},
       eprint = {2104.04824},
 primaryClass = {astro-ph.IM},
       adsurl = {https://ui.adsabs.harvard.edu/abs/2022EPSC...16.1114T}
}

@inproceedings{ricker2016transiting,
  title={The transiting exoplanet survey satellite},
  author={Ricker, George R and Vanderspek, R and Winn, J and Seager, S and Berta-Thompson, Z and Levine, A and Villasenor, J and Latham, D and Charbonneau, D and Holman, M and others},
  booktitle={Space Telescopes and Instrumentation 2016: Optical, Infrared, and Millimeter Wave},
  volume={9904},
  pages={767--784},
  year={2016},
  organization={SPIE}
}

@article{prusti2016gaia,
  title={The gaia mission},
  author={Prusti, Timo and De Bruijne, JHJ and Brown, Anthony GA and Vallenari, Antonella and Babusiaux, C and Bailer-Jones, CAL and Bastian, U and Biermann, M and Evans, Dafydd Wyn and Eyer, L and others},
  journal={A\&A},
  volume={595},
  pages={A1},
  year={2016},
  publisher={EDP sciences}
}

@article{benz2021cheops,
  title={The CHEOPS mission},
  author={Benz, Willy and Broeg, Christopher and Fortier, Andrea and Rando, Nicola and Beck, Thomas and Beck, Mathias and Queloz, Didier and Ehrenreich, David and Maxted, PFL and Isaak, KG and others},
  journal={Experimental Astronomy},
  volume={51},
  number={1},
  pages={109--151},
  year={2021},
  publisher={Springer}
}

@article{rauer2014plato,
  title={The PLATO 2.0 mission},
  author={Rauer, Heike and Catala, Claude and Aerts, Conny and Appourchaux, Thierry and Benz, Willy and Brandeker, Alexis and Christensen-Dalsgaard, J{\o}orgen and Deleuil, Magalie and Gizon, Laurent and Goupil, M-J and others},
  journal={Experimental Astronomy},
  volume={38},
  number={1},
  pages={249--330},
  year={2014},
  publisher={Springer}
}

@article{rauer2025plato,
  title={The PLATO mission},
  author={Rauer, Heike and Aerts, Conny and Cabrera, Juan and Deleuil, Magali and Erikson, Anders and Gizon, Laurent and Goupil, Mariejo and Heras, Ana and Walloschek, Thomas and Lorenzo-Alvarez, Jose and others},
  journal={Experimental Astronomy},
  volume={59},
  number={3},
  pages={26},
  year={2025},
  publisher={Springer}
}

@article{borucki2010kepler,
  title={Kepler planet-detection mission: introduction and first results},
  author={Borucki, William J and Koch, David and Basri, Gibor and Batalha, Natalie and Brown, Timothy and Caldwell, Douglas and Caldwell, John and Christensen-Dalsgaard, J{\o}rgen and Cochran, William D and DeVore, Edna and others},
  journal={Science},
  volume={327},
  number={5968},
  pages={977--980},
  year={2010},
  publisher={American Association for the Advancement of Science}
}

@article{koch2010kepler,
  title={Kepler mission design, realized photometric performance, and early science},
  author={Koch, David G and Borucki, William J and Basri, Gibor and Batalha, Natalie M and Brown, Timothy M and Caldwell, Douglas and Christensen-Dalsgaard, J{\o}rgen and Cochran, William D and DeVore, Edna and Dunham, Edward W and others},
  journal={The Astrophysical Journal Letters},
  volume={713},
  number={2},
  pages={L79},
  year={2010},
  publisher={IOP Publishing}
}

@article{jannsen2024platosim,
  title={PlatoSim: an end-to-end PLATO camera simulator for modelling high-precision space-based photometry},
  author={Jannsen, N and De Ridder, J and Seynaeve, D and Regibo, S and Huygen, R and Royer, P and Paproth, Carsten and Grie{\ss}bach, Denis and Samadi, R and Reese, DR and others},
  journal={Astronomy \& Astrophysics},
  volume={681},
  pages={A18},
  year={2024},
  publisher={EDP Sciences}
}

@article{nascimbeni2025plato,
  title={The PLATO field selection process-II. Characterization of LOPS2, the first long-pointing field},
  author={Nascimbeni, V and Piotto, G and Cabrera, J and Montalto, M and Marinoni, S and Marrese, PM and Aerts, C and Altavilla, G and Benatti, S and B{\"o}rner, A and others},
  journal={Astronomy \& Astrophysics},
  volume={694},
  pages={A313},
  year={2025},
  publisher={EDP Sciences}
}

@ARTICLE{2024CoSka..54b.142D,
       author = {{Deeg}, H.~J. and {Alonso}, R.},
        title = "{Ground-based photometric follow-up for exoplanet detections with the PLATO mission}",
      journal = {Contributions of the Astronomical Observatory Skalnate Pleso},
         year = 2024,
        month = feb,
       volume = {54},
       number = {2},
        pages = {142-149},
          doi = {10.31577/caosp.2024.54.2.142},
archivePrefix = {arXiv},
       eprint = {2401.04503},
 primaryClass = {astro-ph.EP},
       adsurl = {https://ui.adsabs.harvard.edu/abs/2024CoSka..54b.142D}
}

@article{nascimbeni2022plato,
  title={The PLATO field selection process-I. Identification and content of the long-pointing fields},
  author={Nascimbeni, VALERIO and Piotto, G and B{\"o}rner, Anko and Montalto, M and Marrese, PM and Cabrera, J and Marinoni, S and Aerts, C and Altavilla, G and Benatti, S and others},
  journal={Astronomy \& Astrophysics},
  volume={658},
  pages={A31},
  year={2022},
  publisher={EDP Sciences}
}

@article{mantovan2024gaps,
  title={The GAPS programme at TNG-XLIX. TOI-5398, the youngest compact multi-planet system composed of an inner sub-Neptune and an outer warm Saturn},
  author={Mantovan, Giacomo and Malavolta, L and Desidera, S and Zingales, T and Borsato, L and Piotto, G and Maggio, A and Locci, D and Polychroni, D and Turrini, D and others},
  journal={Astronomy \& Astrophysics},
  volume={682},
  pages={A129},
  year={2024},
  publisher={EDP Sciences}
}

@article{tinetti2021ariel,
  title={Ariel: Enabling planetary science across light-years},
  author={Tinetti, Giovanna and Eccleston, Paul and Haswell, Carole and Lagage, Pierre-Olivier and Leconte, J{\'e}r{\'e}my and L{\"u}ftinger, Theresa and Micela, Giusi and Min, Michel and Pilbratt, G{\"o}ran and Puig, Ludovic and others},
  journal={arXiv preprint arXiv:2104.04824},
  year={2021}
}

@article{magrini2023hrmos,
  title={HRMOS white paper: science motivation},
  author={Magrini, Laura and Bensby, Thomas and Brucalassi, Anna and Randich, Sofia and Jeffries, Robin and de Silva, Gayandhi and Skuladottir, Asa and Smiljanic, Rodolfo and Gonzalez, Oscar and Hill, Vanessa and others},
  journal={arXiv preprint arXiv:2312.08270},
  year={2023}
}

@article{kerins2023rosetz,
  title={RoSETZ: Roman Survey of the Earth Transit Zone--a SETI-optimized survey for habitable-zone exoplanets},
  author={Kerins, Eamonn and Awiphan, Supachai and Edmondson, Kathryn and Garrett, Michael and Haqq-Misra, Jacob and Heller, R{\'e}ne and Huston, Macy and Kipping, David and Kopparapu, Ravi and Price, Danny C and others},
  journal={arXiv preprint arXiv:2306.10202},
  year={2023}
}

@article{saba2025mauvesim,
  title={MauveSim: the instrument simulator software for the Blue Skies Space Mauve satellite},
  author={Saba, Arianna and Favata, Fabio and Savini, Giorgio and Tinetti, Giovanna and Bradley, Lawrence and Stotesbury, Ian and Tessenyi, Marcell},
  journal={RAS Techniques and Instruments},
  pages={rzaf045},
  year={2025},
  publisher={Oxford University Press}
}

@article{hey2024confronting,
  title={Confronting sparse Gaia DR3 photometry with TESS for a sample of around 60 000 OBAF-type pulsators},
  author={Hey, Daniel and Aerts, Conny},
  journal={Astronomy \& Astrophysics},
  volume={688},
  pages={A93},
  year={2024},
  publisher={EDP Sciences}
}

@article{bluhm2020precise,
  title={Precise mass and radius of a transiting super-Earth planet orbiting the M dwarf TOI-1235: a planet in the radius gap?},
  author={Bluhm, Paz and Luque, R and Espinoza, N and Pall{\'e}, E and Caballero, Jose A and Dreizler, S and Livingston, JH and Mathur, S and Quirrenbach, A and Stock, S and others},
  journal={Astronomy \& Astrophysics},
  volume={639},
  pages={A132},
  year={2020},
  publisher={EDP Sciences}
}

@article{leonardi2024taste,
  title={TASTE-V. A new ground-based investigation of orbital decay in the ultra-hot Jupiter WASP-12b},
  author={Leonardi, P and Nascimbeni, V and Granata, V and Malavolta, L and Borsato, L and Biazzo, K and Lanza, AF and Desidera, S and Piotto, G and Nardiello, D and others},
  journal={Astronomy \& Astrophysics},
  volume={686},
  pages={A84},
  year={2024},
  publisher={EDP Sciences}
}

@article{biazzo2022gaps,
  title={The GAPS Programme at TNG-XXXV. Fundamental properties of transiting exoplanet host stars},
  author={Biazzo, Katia and D’Orazi, V and Desidera, Silvano and Turrini, Diego and Benatti, Serena and Gratton, Raffaele and Magrini, L and Sozzetti, A and Baratella, M and Bonomo, AS and others},
  journal={Astronomy \& Astrophysics},
  volume={664},
  pages={A161},
  year={2022},
  publisher={EDP Sciences}
}

@article{tonucci2025analysing,
  title={Analysing the flux stability of stellar calibrator candidates with TESS},
  author={Tonucci, Elena and van Kempen, TA and Beaulieu, J-P and Bernard, Lilou},
  journal={Astronomy \& Astrophysics},
  volume={695},
  pages={A277},
  year={2025},
  publisher={EDP Sciences}
}

@Manual{dask,
  title = {Dask: Library for dynamic task scheduling},
  author = {{Dask Development Team}},
  year = {2016},
  url = {http://dask.pydata.org},
}

@article{fabricius2021gaia,
  title={Gaia Early Data Release 3-Catalogue validation},
  author={Fabricius, Claus and Luri, Xavier and Arenou, F and Babusiaux, C and Helmi, A and Muraveva, Tatiana and Reyl{\'e}, C and Spoto, F and Vallenari, A and Antoja, T and others},
  journal={Astronomy \& Astrophysics},
  volume={649},
  pages={A5},
  year={2021},
  publisher={EDP Sciences}
}

@article{stelzer2022great,
  title={The Great Flare of 2021 November 19 on AD Leonis-Simultaneous XMM-Newton and TESS observations},
  author={Stelzer, B and Caramazza, M and Raetz, St and Argiroffi, C and Coffaro, M},
  journal={Astronomy \& Astrophysics},
  volume={667},
  pages={L9},
  year={2022},
  publisher={EDP Sciences}
}

@article{kourniotis2025variability,
  title={Variability of Galactic blue supergiants observed with TESS},
  author={Kourniotis, Michalis and Cidale, Lydia S and Kraus, Michaela and Diaz, Matias A Ruiz and Adam, Aldana Alberici},
  journal={Astronomy \& Astrophysics},
  volume={697},
  pages={A152},
  year={2025},
  publisher={EDP Sciences}
}

@article{calcaferro2024pulsating,
  title={Pulsating hydrogen-deficient white dwarfs and pre-white dwarfs observed with TESS-VI. Asteroseismology of the GW Vir-type central star of the Planetary Nebula NGC 246},
  author={Calcaferro, Leila M and Sowicka, Paulina and Uzundag, Murat and C{\'o}rsico, Alejandro H and Kepler, Souza O and Bell, Keaton J and Althaus, Leandro G and Handler, Gerald and Kawaler, Steven D and Werner, Klaus},
  journal={Astronomy \& Astrophysics},
  volume={686},
  pages={A140},
  year={2024},
  publisher={EDP Sciences}
}

@article{castro2024toi,
  title={TOI-5005 b: A super-Neptune in the savanna near the ridge},
  author={Castro-Gonz{\'a}lez, A and Lillo-Box, J and Armstrong, DJ and Acu{\~n}a, L and Aguichine, A and Bourrier, V and Gandhi, S and Sousa, SG and Delgado-Mena, E and Moya, A and others},
  journal={Astronomy \& Astrophysics},
  volume={691},
  pages={A233},
  year={2024},
  publisher={EDP Sciences}
}

@article{bentley1975multidimensional,
  title={Multidimensional binary search trees used for associative searching},
  author={Bentley, Jon Louis},
  journal={Communications of the ACM},
  volume={18},
  number={9},
  pages={509--517},
  year={1975},
  publisher={ACM New York, NY, USA}
}

@InProceedings{ mckinney-proc-scipy-2010,
  author    = { {W}es {M}c{K}inney },
  title     = { {D}ata {S}tructures for {S}tatistical {C}omputing in {P}ython },
  booktitle = { {P}roceedings of the 9th {P}ython in {S}cience {C}onference },
  pages     = { 56 - 61 },
  year      = { 2010 },
  editor    = { {S}t\'efan van der {W}alt and {J}arrod {M}illman },
  doi       = { 10.25080/Majora-92bf1922-00a }
}

@article{virtanen2020scipy,
  author  = {Virtanen, Pauli and others},
  title   = {SciPy 1.0: Fundamental Algorithms for Scientific Computing in Python},
  journal = {Nature Methods},
  volume  = {17},
  pages   = {261--272},
  year    = {2020},
  doi     = {10.1038/s41592-019-0686-2}
}

@article{harris2020array,
  title={Array programming with NumPy},
  author={Harris, Charles R and Millman, K Jarrod and Van Der Walt, St{\'e}fan J and Gommers, Ralf and Virtanen, Pauli and Cournapeau, David and Wieser, Eric and Taylor, Julian and Berg, Sebastian and Smith, Nathaniel J and others},
  journal={nature},
  volume={585},
  number={7825},
  pages={357--362},
  year={2020},
  publisher={Nature Publishing Group UK London}
}

@article{torra2021gaia,
  title={Gaia Early Data Release 3-Building the Gaia DR3 source list--Cross-match of Gaia observations},
  author={Torra, Ferran and Casta{\~n}eda, J and Fabricius, Claus and Lindegren, L and Clotet, M and Gonz{\'a}lez-Vidal, JJ and Bartolom{\'e}, S and Bastian, Ulrich and Bernet, M and Biermann, Michael and others},
  journal={Astronomy \& Astrophysics},
  volume={649},
  pages={A10},
  year={2021},
  publisher={EDP Sciences}
}

@article{babusiaux2023gaia,
  title={Gaia data release 3-catalogue validation},
  author={Babusiaux, C and Fabricius, C and Khanna, S and Muraveva, Tatiana and Reyl{\'e}, C and Spoto, F and Vallenari, Antonella and Luri, X and Arenou, F and Alvarez, MA and others},
  journal={Astronomy \& Astrophysics},
  volume={674},
  pages={A32},
  year={2023},
  publisher={EDP Sciences}
}

@inproceedings{mcglynn1998skyview,
  title={Skyview: The multi-wavelength sky on the interne},
  author={McGlynn, T and Scollick, K and White, N},
  booktitle={Symposium-International Astronomical Union},
  volume={179},
  pages={465--466},
  year={1998},
  organization={Cambridge University Press}
}

@article{lasker1990guide,
  title={The guide star catalog. i-astronomical foundations and image processing},
  author={Lasker, Barry M and Sturch, Conrad R and McLean, Brian J and Russell, Jane L and Jenkner, Helmut and Shara, Michael M},
  journal={Astronomical Journal (ISSN 0004-6256), vol. 99, June 1990, p. 2019-2058, 2173-2178.},
  volume={99},
  pages={2019--2058},
  year={1990}
}

@article{feinstein2019eleanor,
  title={eleanor: An open-source tool for extracting light curves from the TESS Full-Frame Images},
  author={Feinstein, Adina D and Montet, Benjamin T and Foreman-Mackey, Daniel and Bedell, Megan E and Saunders, Nicholas and Bean, Jacob L and Christiansen, Jessie L and Hedges, Christina and Luger, Rodrigo and Scolnic, Daniel and others},
  journal={Publications of the Astronomical Society of the Pacific},
  volume={131},
  number={1003},
  pages={094502},
  year={2019},
  publisher={The Astronomical Society of the Pacific}
}

@article{han2023tess,
  title={TESS--Gaia Light Curve: A PSF-based TESS FFI Light-curve Product},
  author={Han, Te and Brandt, Timothy D},
  journal={The Astronomical Journal},
  volume={165},
  number={2},
  pages={71},
  year={2023},
  publisher={The American Astronomical Society}
}

@article{han2025hundreds,
  title={Hundreds of TESS exoplanets might be larger than we thought},
  author={Han, Te and Robertson, Paul and Brandt, Timothy D and Kanodia, Shubham and Ca{\~n}as, Caleb and Shporer, Avi and Ricker, George and Beard, Corey},
  journal={The Astrophysical Journal Letters},
  volume={988},
  number={1},
  pages={L4},
  year={2025},
  publisher={The American Astronomical Society}
}

@article{edwards2022ariel,
  title={The Ariel target list: The impact of TESS and the potential for characterizing multiple planets within a system},
  author={Edwards, Billy and Tinetti, Giovanna},
  journal={The Astronomical Journal},
  volume={164},
  number={1},
  pages={15},
  year={2022},
  publisher={IOP Publishing}
}

@article{mugnai2020arielrad,
  title={ArielRad: the Ariel radiometric model},
  author={Mugnai, Lorenzo V and Pascale, Enzo and Edwards, Billy and Papageorgiou, Andreas and Sarkar, Subhajit},
  journal={Experimental Astronomy},
  volume={50},
  number={2},
  pages={303--328},
  year={2020},
  publisher={Springer}
}

@article{edwards2019updated,
  title={An updated study of potential targets for Ariel},
  author={Edwards, Billy and Mugnai, Lorenzo and Tinetti, Giovanna and Pascale, Enzo and Sarkar, Subhajit},
  journal={The Astronomical Journal},
  volume={157},
  number={6},
  pages={242},
  year={2019},
  publisher={American Astronomical Society}
}

@techreport{nascimbeni2017ariel,
  author      = {Nascimbeni, V. and Piotto, G.},
  title       = {Contaminant Analysis for {ARIEL}},
  institution = {Istituto Nazionale di Astrofisica (INAF)},
  year        = {2017},
  number      = {ARIEL-INAF-SCI-TN-004},
  month       = {mar},
  type        = {Technical Report},
  note        = {ARIEL Space Mission Assessment Phase},
  url         = {https://wordpress.com}
}

@article{mugnai2025exosim,
  title={ExoSim 2: the new exoplanet observation simulator applied to the Ariel space mission},
  author={Mugnai, Lorenzo V and Bocchieri, Andrea and Pascale, Enzo and Lorenzani, Andrea and Papageorgiou, Andreas},
  journal={Experimental Astronomy},
  volume={59},
  number={1},
  pages={9},
  year={2025},
  publisher={Springer}
}

@article{allard2012models,
  title={Models of very-low-mass stars, brown dwarfs and exoplanets},
  author={Allard, France and Homeier, Derek and Freytag, Bernd},
  journal={Philosophical transactions. Series A, Mathematical, physical, and engineering sciences},
  volume={370},
  number={1968},
  pages={2765},
  year={2012}
}

@INPROCEEDINGS{2003IAUS..210P.A20C,
       author = {{Castelli}, F. and {Kurucz}, R.~L.},
        title = "{New Grids of ATLAS9 Model Atmospheres}",
    booktitle = {Modelling of Stellar Atmospheres},
         year = 2003,
       editor = {{Piskunov}, N. and {Weiss}, W.~W. and {Gray}, D.~F.},
       series = {IAU Symposium},
       volume = {210},
        month = jan,
        pages = {A20},
          doi = {10.48550/arXiv.astro-ph/0405087},
archivePrefix = {arXiv},
       eprint = {astro-ph/0405087},
 primaryClass = {astro-ph},
       adsurl = {https://ui.adsabs.harvard.edu/abs/2003IAUS..210P.A20C}
}

@article{lanz2003grid,
  title={A grid of non-LTE line-blanketed model atmospheres of O-type stars},
  author={Lanz, Thierry and Hubeny, Ivan},
  journal={The Astrophysical Journal Supplement Series},
  volume={146},
  number={2},
  pages={417--441},
  year={2003}
}

@article{lanz2007grid,
  title={A grid of NLTE line-blanketed model atmospheres of early B-type stars},
  author={Lanz, Thierry and Hubeny, Ivan},
  journal={The Astrophysical Journal Supplement Series},
  volume={169},
  number={1},
  pages={83--104},
  year={2007}
}

@article{riello2021gaia,
  title={Gaia early data release 3: photometric content and validation},
  author={Riello, Marco and De Angeli, F and Evans, DW and Montegriffo, P and Carrasco, JM and Busso, G and Palaversa, Lovro and Burgess, PW and Diener, C and Davidson, M and others},
  journal={Astronomy \& astrophysics},
  volume={649},
  pages={A3},
  year={2021},
  publisher={EDP Sciences}
}

@misc{gaiaDR3photometricrelations,
  author       = {{Gaia Collaboration}},
  title        = {Gaia Data Release 3: Photometric relations},
  year         = {2022},
  howpublished = {\url{https://gea.esac.esa.int/archive/documentation/GDR3/Data_processing/chap_cu5pho/cu5pho_sec_photSystem/cu5pho_ssec_photRelations.html}},
  note         = {Gaia DR3 documentation}
}

@article{fritsch1980monotone,
  title={Monotone piecewise cubic interpolation},
  author={Fritsch, Frederick N and Carlson, Ralph E},
  journal={SIAM Journal on Numerical Analysis},
  volume={17},
  number={2},
  pages={238--246},
  year={1980},
  publisher={SIAM}
}

@inproceedings{rodrigo2020svo,
  title={The SVO filter profile service},
  author={Rodrigo, Carlos and Solano, Enrique},
  booktitle={XIV. 0 Scientific Meeting (virtual) of the Spanish Astronomical Society},
  pages={182},
  year={2020}
}

@article{brown2020gaia,
  title={Gaia Early Data Release 3: Summary of the contents and survey properties},
  author={Brown, AG and Vallenari, Antonella and Prusti, Timo and De Bruijne, Jos HJ and Babusiaux, Carine and Biermann, Michael and Creevey, Orlagh L and Evans, Dafydd Wyn and Eyer, Laurent and Hutton, A and others},
  journal={arXiv preprint arXiv:2012.01533},
  year={2020}
}

@ARTICLE{2015ApJ...805...16C,
       author = {{Ciardi}, David R. and {Beichman}, Charles A. and {Horch}, Elliott P. and {Howell}, Steve B.},
        title = "{Understanding the Effects of Stellar Multiplicity on the Derived Planet Radii from Transit Surveys: Implications for Kepler, K2, and TESS}",
      journal = {\apj},
         year = 2015,
       volume = {805},
       number = {1},
          eid = {16},
        pages = {16},
          doi = {10.1088/0004-637X/805/1/16}
}

@ARTICLE{2020ApJ...898...47F,
       author = {{Furlan}, E. and {Howell}, S.~B.},
        title = "{Unresolved Binary Exoplanet Host Stars Fit as Single Stars: Effects on the Stellar Parameters}",
      journal = {\apj},
         year = 2020,
       volume = {898},
       number = {1},
          eid = {47},
        pages = {47},
          doi = {10.3847/1538-4357/ab9c9c}
}

@ARTICLE{2022ApJ...935..141S,
       author = {{Sullivan}, Kendall and {Kraus}, Adam L. and {Mann}, Andrew W.},
        title = "{Revising Properties of Planet-Host Binary Systems. I. Methods and Pilot Study}",
      journal = {\apj},
         year = 2022,
       volume = {935},
       number = {2},
          eid = {141},
        pages = {141},
          doi = {10.3847/1538-4357/ac7be9}
}

@ARTICLE{2022AJ....164..138S,
       author = {{Sullivan}, Kendall and {Kraus}, Adam L.},
        title = "{Revising Properties of Planet-Host Binary Systems. II. Apparent Near-Earth-analog Planets in Binaries Are Often Sub-Neptunes}",
      journal = {\aj},
         year = 2022,
       volume = {164},
       number = {4},
          eid = {138},
        pages = {138},
          doi = {10.3847/1538-3881/ac89ed}
}

@ARTICLE{2021AJ....162...75L,
       author = {{Lester}, Kathryn V. and {Matson}, Rachel A. and {Howell}, Steve B. and {Furlan}, Elise and {Gnilka}, Crystal L. and {Scott}, Nicholas J. and {Ciardi}, David R. and {Everett}, Mark E. and {Hartman}, Zachary D. and {Hirsch}, Lea A.},
        title = "{Speckle Observations of TESS Exoplanet Host Stars. II. Stellar Companions at 1-1000 au and Implications for Small Planet Detection}",
      journal = {\aj},
         year = 2021,
       volume = {162},
       number = {2},
          eid = {75},
        pages = {75},
          doi = {10.3847/1538-3881/ac0d06}
}





\bsp	
\label{lastpage}
\end{document}